%% file: cuore-ppnp.tex
\documentclass[1p]{elsarticle}
\usepackage{lineno,hyperref}
\usepackage{comment}
\usepackage{graphicx}% Include figure files
\usepackage{soul}
\usepackage{newtxtext,newtxmath,amsmath}

\modulolinenumbers[1]

\usepackage{lipsum}

\journal{Progress in Particle and Nuclear Physics}

\newcommand{\twonu}{\ensuremath{2\nu\beta\beta}}
\newcommand{\zeronu}{\ensuremath{0\nu\beta\beta}}

\usepackage{siunitx}
\usepackage[final, inline, nomargin]{fixme}
\fxsetup{theme=color, mode=multiuser}
\FXRegisterAuthor{pts}{1}{\color{brown}Notes to self (Pranava)}

\usepackage{xcolor}
\definecolor{cerulean}{rgb}{0.0, 0.48, 0.65}
\definecolor{amaranth}{rgb}{0.9, 0.17, 0.31}

\begin{document}

\begin{frontmatter}

%\title{CUORE opens the door to ton scale cryogenics experiment\tnoteref{mytitlenote}}
%\tnotetext[mytitlenote]{This is a tentative title, changes could be applied during the article writing phase.}

\title{CUORE Opens the Door to Tonne-scale Cryogenics Experiments}

% Updated on Friday, 6 May 2021 
%\input{cuore-authors.tex}

\input{cuore-authors}

\begin{abstract}
The past few decades have seen major developments in the design and operation of cryogenic particle detectors. This technology offers an extremely good energy resolution --- comparable to semiconductor detectors --- and a wide choice of target materials, making low temperature calorimetric detectors ideal for a variety of particle physics applications.
Rare event searches have continued to require ever greater exposures, which has driven them to ever larger cryogenic detectors, with the CUORE experiment being the first to reach a tonne-scale, mK-cooled, experimental mass.
CUORE, designed to search for neutrinoless double beta decay, has been operational since 2017 at a temperature of about 10 mK. 
This result has been attained by the use of an unprecedentedly large cryogenic infrastructure called the CUORE cryostat: conceived, designed and commissioned for this purpose.

In this article the main characteristics and features of the cryogenic facility developed for the CUORE experiment are highlighted.
A brief introduction of the evolution of the field and of the past cryogenic facilities are given. 
The motivation behind the design and development of the CUORE cryogenic facility is detailed as are the steps taken toward realization, commissioning, and operation of the CUORE cryostat. 
The major challenges overcome by the collaboration and the solutions implemented throughout the building of the cryogenic facility will be discussed along with the potential improvements for future facilities.

The success of CUORE has opened the door to a new generation of large-scale cryogenic facilities in numerous fields of science.
Broader implications of the incredible feat achieved by the CUORE collaboration on the future cryogenic facilities in various fields ranging from neutrino and dark matter experiments to quantum computing will be examined.
\end{abstract}

%\begin{keyword}
%\texttt{elsarticle.cls}\sep \LaTeX\sep Elsevier \sep template
%\MSC[2010] 00-01\sep  99-00
%\end{keyword}

\end{frontmatter}

% \linenumbers

\section{Introduction}\label{sec:intro}
Cryogenics is a branch of physics and engineering that studies the techniques to reach temperatures well below room temperature and the behavior of materials under such conditions. 
In the last two decades one of the driving forces behind the advancement of cryogenics at mK temperatures has been the development of low temperature detectors to explore new frontiers in physics and astrophysics.~\ptsnote{Need a citation here. Suggested by Brad.}
However, these advancements have also benefited other research fields including solid state physics, materials research, nuclear physics, and the emerging interdisciplinary field of quantum information science (QIS). 
In fact the recent advances towards quantum sensors, quantum communications, and quantum computing relied heavily on cryogenics because most quantum devices operate near absolute zero. 

In this article the biggest mK-scale cryogenics facility ever built for the study of rare events in particle physics is presented. 
This paper is organized as follows:
Section~\ref{sec:CryoDet} gives an introduction to very low temperature detectors and briefly describes the impetus behind the development of large cryogenic detectors for rare event searches.
Section~\ref{sec:DilRef} presents the principles of a dilution refrigerator while Section~\ref{sec:LargeDRs} provides some historical context to the development of dilution refrigerators.
Section~\ref{sec:CUORE} briefly describes neutrinoless double beta decay and the experimental strategy of CUORE to search for this rare process.
Section~\ref{sec:cuore_cryostat} provides a detailed description of the CUORE cryogenic facility.
Section~\ref{sec:commissioning} outlines the commissioning and performance of the CUORE cryostat.
Section~\ref{sec:future} examines the potential and future applications of large cryogenic facilities based on the success of the CUORE cryogenic facility and section~\ref{sec:conclusion} provides concluding remarks.

\section{Cryogenic detectors}\label{sec:CryoDet}

The historic development of cryogenic detectors was prompted in the 1990's --- primarily within the astroparticle physics community --- by the need for particle detectors capable of detecting small energy depositions with good energy resolution ~\cite{Enns2005,twerenbold1996cryogenic,Pretzl2020}. These detectors are basically calorimeters in which the  energy deposited by an interacting radiation converts into a temperature rise that is measured by means of a temperature sensor. 
%The need for detectors capable of identifying spectral signatures in the keV--MeV range was driven by several neutrino and dark matter experiments, opening the way to a completely new approach in particle detection. 
The fundamental need to use low temperature calorimetric detectors (LTDs) is based on the fact that, in a conventional detector, most of the energy released by radiation interacting with a medium is split among multiple channels like scintillation, ionization, and heat.
The energy resolution of such a detector is statistically limited by the number of the excitation quanta released per unit energy into the channels used for detection.
The typical quanta of energies of most widely used detectors, like scintillator and semiconductor detectors, are $\mathcal{O}$(1-10 eV)~\cite{knoll2010radiation,sadoulet1988cryogenic}.
On the other hand, in LTDs the energy deposition in the absorber is primarily in the form of phonons whose energy quanta vary from meV at 300 K down to $\mu$eV at 10 mK. 
%The actual mechanism that the limits the energy resolution of an LTD is based on the idea of multiple measurements of system energy under equilibrium and in principle can be arbitrarily small over a large measurement period at low temperature.
In practice the limits on energy resolution comes from the thermal equilibrium with the heat sink and is based on the heat capacity as
\begin{equation}
    \Delta E \propto \sqrt{k C T^2},
\end{equation}
where k is the Boltzmann constant, C is the heat capacity and T the sink temperature.
% Detecting the thermal signal provides a further advantage as the energy of the carrier is very small, improving the intrinsic resolution.
% To obtain good energy resolution, it is thus favourable to have most of the energy deposited in small quanta of energy into a detectable channel.
% On the other hand, in LTDs the energy deposition in the absorber is primarily in the form of phonons whose energy quanta vary from 10$^{-3}$ e\si{\volt}  at 300\,K down to $\mathcal{O}$($\mu$eV) at 10\,mK.
% However, to measure such a signal and to minimize thermal noise the detector has be operated at extremely low temperatures.
Thus to measure the deposited energy and minimize thermal noise, the detector must be operated at extremely low temperatures.
Further, the heat capacity of dielectric materials follows the Debye law which scales as $T^{3}$ and thus is extremely small at low temperatures. 
Minimizing the heat capacity in turn maximizes the temperature rise for a unit energy deposition in the absorber allowing for improved detection.
%Consequently, it is crucial to maintain the detector absorber at very low temperatures of $\mathcal{O}$(m$\si{\kelvin}$). 

These detectors, in addition to having excellent energy resolution and low energy thresholds, offer the possibility of using a wide selection of materials for the absorber. 
All these features make cryogenic detectors ideal for a wide range of applications. 
The development of LTDs, in particular for rare event astroparticle physics searches, has significantly aided the development of better performing and larger cryogenic infrastructures~\cite{Pirro2017,gorla2006cuoricino}. 
The CUORE experiment is the culmination of decades of research in cryogenic techniques and its success has provided the stimulus for the next generation of large-scale cryogenic facilities applicable to several fields of science.

\subsection{Search for rare events}\label{sec:rare}

Although calorimetric detectors have been used for a long period of time~\cite{langley,ellis1927average,simon1935application}, the use of cryogenic detectors in elementary particle physics experiments was proposed in early 1970's~\cite{lubkin1974gentle,niinikoski1989cryogenic}. 
Further development in cryogenic detectors was motivated by the proposals put forward in the early 1980's to perform rare event searches and X-ray spectroscopy measurements.

In 1984, E.~Fiorini and T.~O.~Niinikoski proposed the use of cryogenic calorimeters to improve the sensitivity of double beta decay measurements~\cite{fiorini1984low}. 
In the same year, A.~Drukier and L.~Stodolsky proposed the use of superconducting micro-grains to detect neutrinos scattered coherently off nuclei~\cite{Drukier84} with high cross-section. 
Following this idea, M.~W.~Goodman and E.~Witten~\cite{Goodman85} proposed to use cryogenic detectors for detecting dark matter~(DM) candidates.
Contemporaneously, in 1984 D.~McCammon, S.~Moseley~\cite{mccammon1984experimental,moseley1984thermal} and collaborators have proposed the use of thermal X-ray microcalorimeters in X-ray astronomy. 
The R\&D work following this proposal had pushed the development of the cryogenic calorimeter for rare event searches.

In the 1990's the field underwent significant developments propelled by the neutrinoless double beta decay~(\zeronu)~\cite{CUORE_PRECURSOR} and DM~\cite{Proebst95,deBellefon96, Barnes:1996db,cebrian2002rosebud} search experiments. 
Starting in the early 1990's Fiorini and collaborators started developing LTDs based on TeO$_2$ crystals to search for neutrinoless double beta decay. 
In few years the TeO$_2$ LTDs grew from few grams to kg scale~\cite{CUORE_PRECURSOR}. 
Concurrent progress was made in developing phonon-mediated detectors for DM search experiments.
Following the pioneering works by F.~Avignone and collaborators~\cite{Ahlen:1987mn} and O.~Cremonesi~\cite{Cremonesi:1986ij} using Ge detectors, early 1990's several DM experiments were developed using cryogenic detectors. 
%These developments were driven by the need for improvements in the performance of energy resolution and threshold barely reachable using conventional detectors. 
The CDMS~\cite{CDMS2000}, CRESST~\cite{angloher2002limits}, and EDELWEISS~\cite{EDELWEISS2001} collaborations developed the first prototypes of cryogenic detectors and started to explore the WIMP~(Weakly Interacting Massive Particle) DM parameter space~\cite{Roszkowski:2017nbc}.
%In the '90s the field underwent a great development with absorber crystals of increasing mass of the 100\,g-1\,kg scale for most of these projects.}

The experiments, both in the field of DM search and double beta decay, are considered ``rare event'' search projects. 
They are characterized by the need to simultaneously achieve low backgrounds, excellent energy resolution, and large exposure. 
These needs pushed the experimental collaborations to develop larger masses for the absorber crystals to improve the number of nuclei under observation. 
While the focus of detector R\&D in the double beta decay experiments has been to improve detector resolution, crucial for peak identification~\cite{biassoni2020search}, the DM experiments primarily aimed at lowering threshold to allow for a precise measurement of the quasi-exponential DM spectrum.
DM experiments have further developed double readout detectors where the impressive resolution of LTDs coupled with an ionization (CDMS, EDELWEISS) or a scintillation (CRESST) signal allows for radiation discrimination and background rejection. 
In order to achieve  the individual goals of these experiments, a significant effort was seen in the development of large customized cryostats capable of providing not only extremely low operational temperatures, but also a low noise and low background environment.

\section{Dilution refrigerators} 
\label{sec:DilRef}

As discussed in Section~\ref{sec:intro}, cryogenic detectors must be operated at temperatures of $\mathcal{O}$(10\,mK) to provide optimal performance for single particle radiation detection. To date the only cryogenic system capable of reaching and maintaining these stable working temperatures over long periods of time is dilution refrigerator~(DR). 
A DR is a cryogenic device whose working principle relies on the properties of a mixture of two isotopes of helium: $^3$He and $^4$He. 
The idea was originally proposed by H.~London in 1951~\cite{London51} and, about ten years later, H.~London, G.R.~Clarke, and E.~Mendoza published a detailed concept for realizing such a cooling system~\cite{London1962}. 
The first DR prototype was successfully operated in 1965 by P.~Das, R.~B.~de Ouboter, and K.~W.~Taconis~\cite{Taconis65}. 
Steady improvements followed, leading to a lowest stable temperature of 2\,mK obtained by G. Frossati and co-workers in 1978 at the University of Leiden~\cite{frossati78}.
% and of 1.75 mK obtained  by in 1998 \cite{Cousins99}. 
DR technology has since seen continuous progress with improved performance, stability and reliability. 
DR have also been commercially available~\cite{Oxford} since the '60s which allowed for numerous advances in studies at cryogenic temperatures.

\begin{figure}[!t]
    \centering\includegraphics[width=\textwidth]{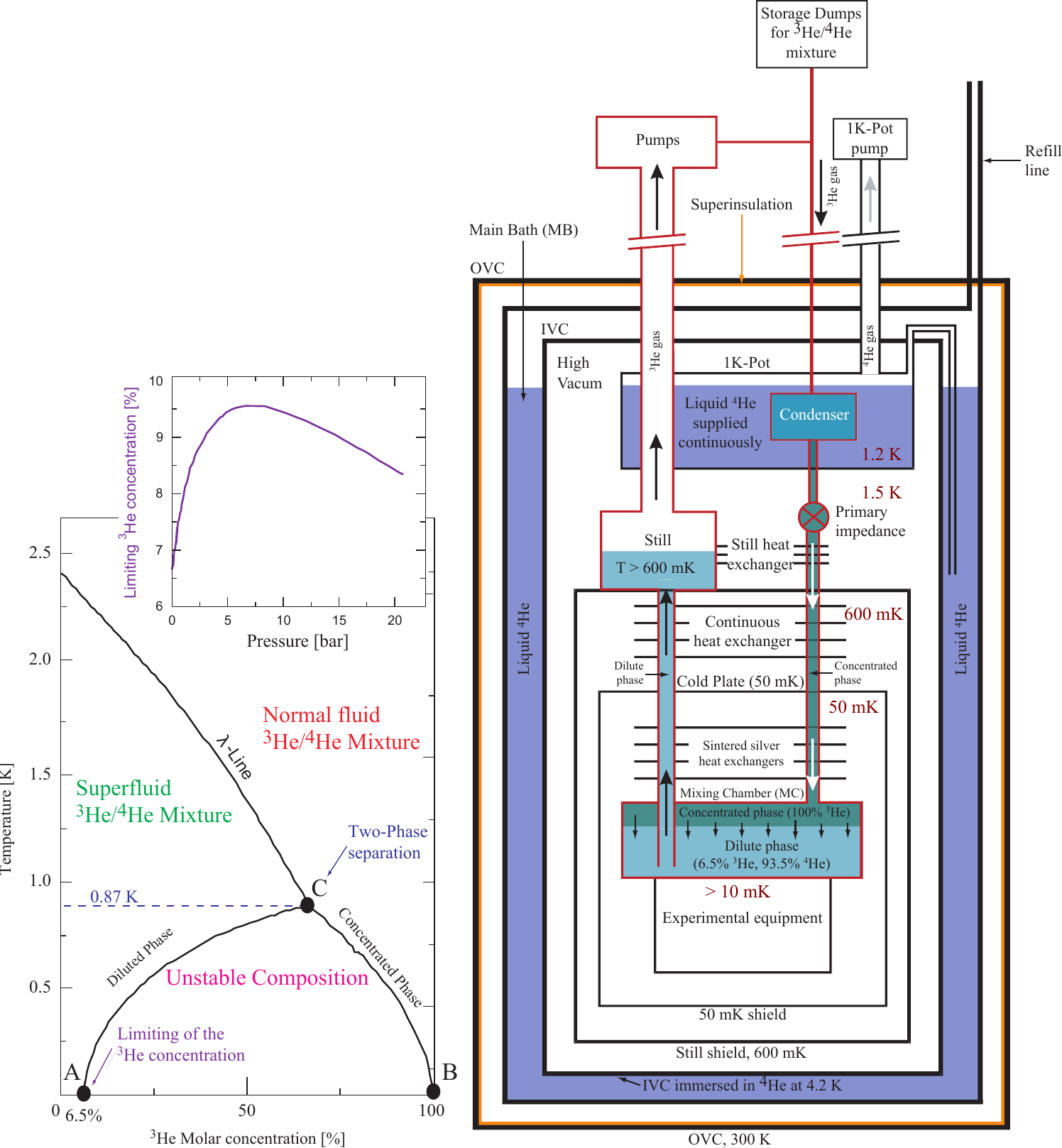}
    \caption{Left: Phase diagram of $^3$He/$^4$He mixture as function of the temperature in saturated vapor pressure. Right: Schematic diagram of a conventional dilution refrigerator.}
    \label{fig:DRbasic}
\end{figure}

%\subsection{Basic of Dilution Refrigerators}
\subsection{\textsuperscript{3}He-\textsuperscript{4}He mixture properties}
To understand the working principle of DRs it is crucial to understand the properties of $^3$He/$^4$He mixture. 
Figure~\ref{fig:DRbasic} on the left shows the relation between the $^3$He molar concentration and temperature for liquid $^3$He/$^4$He mixture at saturated vapor pressure. 
Liquid $^4$He acts as a super-fluid --- a state of matter in which matter behaves like a fluid with zero viscosity --- at $T = 2.177$\,K, while $^3$He does not undergo any phase transition down to few mK. 
However, the temperature of the super-fluid phase transition of the (Bose) liquid $^4$He decreases if this is diluted into the (Fermi) liquid $^3$He. 
When cooled below $T_{\lambda} = 867$\,mK, the mixture undergoes a spontaneous phase separation giving rise to a $^3$He-rich phase (concentrated phase, $H_c$) and a $^3$He-poor phase (diluted phase, $^4$He-rich, $H_d$). 
The two phases are maintained in liquid-vapor form. 
Since there is a boundary between both phases, extra energy is required for particles to transition between phases. 
The lighter $^3$He rich fraction floats on top of the heavier $^4$He rich fraction.

Since the $^3$He vapor pressure is much higher than the $^4$He vapor pressure, by pumping on the $^4$He-rich phase it is possible to remove mostly $^3$He, thus disturbing the equilibrium. 
The equilibrium can be restored by the crossing of $^3$He from the $^3$He-rich side to the $^4$He-rich side. 
The energy needed for the $^3$He to cross the phase boundary is supplied by the surrounding chamber~(typically the so called "mixing chamber") in the form of heat, further cooling the mixture. 
Finally, the atoms lost by the $^3$He-rich phase are replenished by a constantly circulating flow of $^3$He (recirculating dilution refrigeration). 
However, since the enthalpy of $^3$He in the dilute phase is larger than its enthalpy in the concentrated phase, at constant pressure, a constant $n_3$ moles of $^3$He can be made to pass from the concentrated to the diluted phase to obtain the following cooling power:

\begin{equation}\label{eq:appx:hmixing}
\dot{Q}=\dot{n}_3\left[H_d(T)-H_c(T)\right]\simeq 84\,\dot{n}_3\,T^2.
\end{equation}

\noindent Typical values of $\dot{Q}$ for a DR working at $\sim 10$ \,mK are of the order of few $\mu$W.

\subsection{Basics of dilution refrigerators}\label{sec:Bas_DR}
To exploit the cooling properties of the $^3$He/$^4$He, a preliminary cooling is needed in order to reach $T= T_{\lambda}$, where the phase separation takes place. 
In a dilution refrigerator this can be achieved by means of a sequence of steps.

The mixture $^3$He/$^4$He is cooled down to the temperature of 4.2\,K by dipping the dilution unit in a bath (Main Bath, MB) of liquid $^4$He (L$^4$He or commonly LHe). 
The MB is thermally isolated from room temperature by a dewar called Outer Vacuum Chamber~(OVC). 
The second step is realized by an evaporation refrigerator usually called the "1K-pot". 
The 1K-pot is thermally isolated from the 4.2\,K stage by the Inner Vacuum Chamber~(IVC) and is filled by means of a capillary from in the MB. 
Pumping the vapor above the liquid, the 1K-pot temperature decreases to 1.2\,K, according to the Clausius-Clapeyron relation. 

At this point the temperature is low enough to start the dilution refrigeration operation. 
The core of a DR essentially consists of three elements: the Mixing Chamber~(MC), the Still and the Heat Exchangers~(HEX). 
At the temperature of 1.2\,K the mixture passes through an impedance undergoing Joule-Thomson~(JT) expansion and starts to condense in the dilution unit filling the MC first, and then the Still. 
At this stage the liquid mixture is homogeneous. 
In order to get phase separation, the temperature must be reduced to below $T_{\lambda} = 867$\,mK, the tricritical point. 
The Still is the first part with a temperature below 1.2\,K. 
When pumped, the mixture undergoes an expansion and its temperature cools, according to the gas equation of state and when $T < T_{\lambda} $ dilution begins. 
The Still is kept at temperature of $\sim$600-700\,mK and its task is to cool the incoming $^3$He before it enters the HEXs and the MC. 
Finally, the dilution process continues and the cooling proceeds, as explained in the previous section, until the heat extraction rate $\dot{Q}=84\,\dot{n}_3\,T^2$ is balanced by the sum of the various heat inputs present in the system. 
In a typical operating configuration the IVC thermal shield, immersed in the LHe MB, stays at 4.2\,K, the Still shield at $\sim$600\,mK, the HEX shield at 50 mK and the mixing chamber reaches the base temperature (between 5 and 15\,mK for most commercial DRs).

A complete discussion of the working principles and features of DRs is beyond the scope of this work. 
Detailed discussions on the topic can be found in \cite{Betts1989, Lounasmaa1974, Pobell2007, White2001, Enns2005}.

%\textcolor{red}{Cryo strcture, liquid bath, 1K-pot, etc.}

%\begin{figure}[!t]
%    \centering\includegraphics[width=0.7\textwidth]{images/pulsetube.pdf}
%    \caption{A schematic representation of a Pulse Tube Refrigerator. From left to right it consists of a compressor (or rotary valve), an aftercooler (AC), a regenerator, a cold heat exchanger (CHX), a pulse tube (tube), a hot heat exchanger (HHX), an orifice (O), and a buffer}
%    \label{fig:pulsetube}
%\end{figure}

%\textcolor{red}{Cryo strcture, liquid bath, 1K pot, etc.}

\begin{figure}[!t]
    \centering\includegraphics[width=0.7\textwidth]{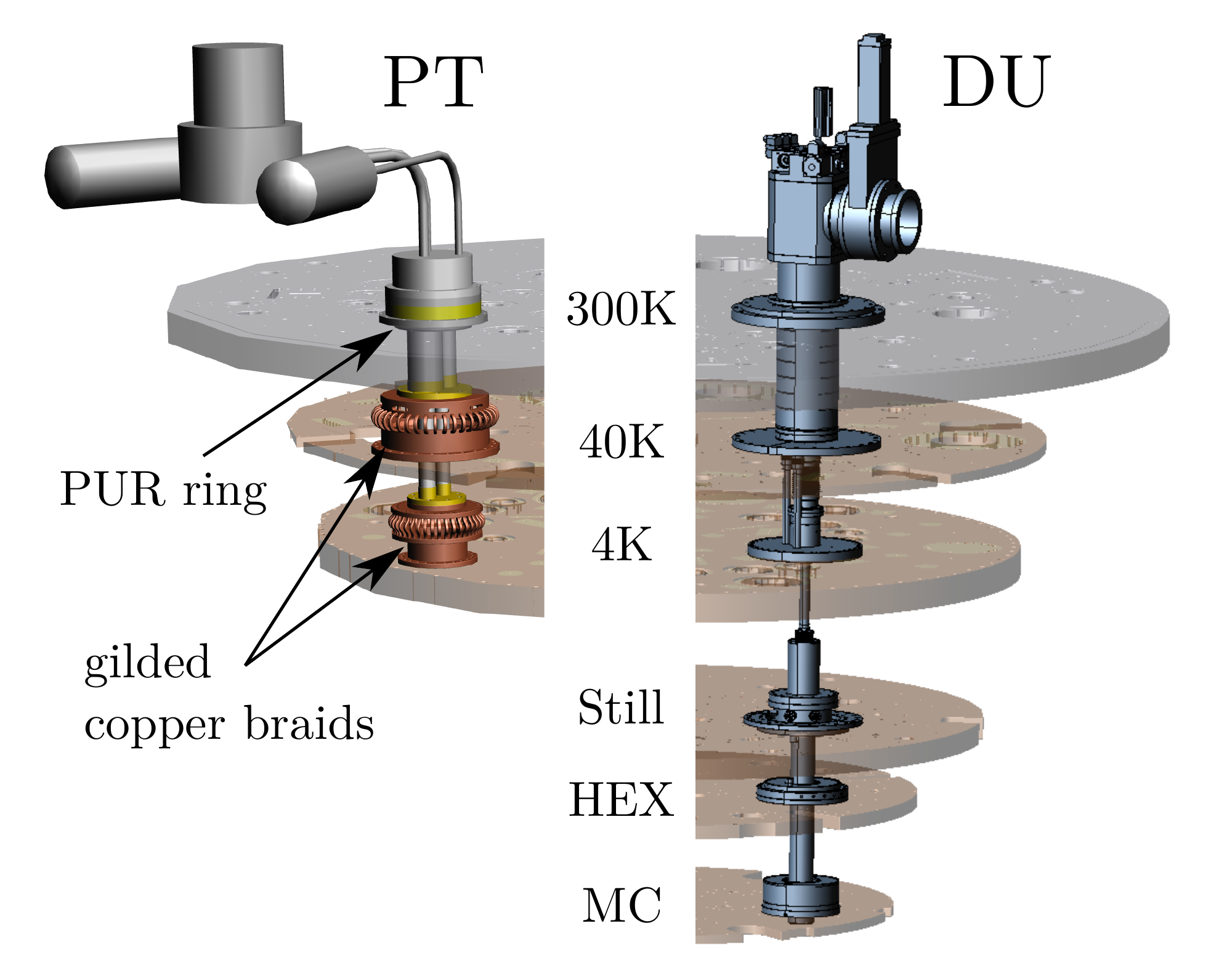}
    \caption{Rendering showing various cooling stages of the CUORE cryostat. (Left) One of the five Pulse Tubes (PTs). (Right) The dilution unit (DU).}
    \label{fig:pulsetube}
\end{figure}

\subsection{Liquid helium bath-free DRs}

One of the main complications of DR-based cryostats is the use of liquid helium bath to supply the 4.2\,K and 1.2\,K stages (described in \ref{sec:Bas_DR}) and, in some apparatus, of liquid nitrogen to supply a thermal stage at 77\,K. 
The use of cryogenic liquids introduces complexity and cost in the procurement~(particularly for liquid helium), shorter duty cycle~(due to the periodic bath refills), and system instability due to the presence of cold liquid whose level is not stable during the operations.

Since the '90s, several attempts have been made to replace the liquid bath in DRs with cold heads~\cite{Pari90, Uhlig93} which was strongly boosted as Pulse Tube~(PTs) cryocoolers became available. 
The idea behind this technique was introduced in 1964 by W.E.~Gifford and R.C.~Longworth~\cite{Gifford1964}. 
Temperatures below 2\,K was achieved in 1990~\cite{Matsubara1994} with a three-stage pulse tube after significant improvements,. 
Currently double stage PTs are commercially available from several companies and can provide cooling powers up to 2.0\,W @ 4.2\,K and 55\,W @45K  (Cryomech PT420) \cite{cryomech, janis}. 
In a PT, a compressor with rotating valve generates a smooth periodic, close to adiabatic, pressure variation and displacement of the working gas in a thin-walled tube with heat exchangers at both ends. 
The latter unit is the actual pulse tube, after which the cryocooler is named. 
The second part consists of a regenerator --- containing a porous magnetic material of high heat capacity --- as a heat reservoir which is connected at its cold end to the PT. 
A detailed description of the working principles of PTs can be found in \cite{Pobell2007,Enns2005}.
The main advantage of this technique compared to other closed-circle refrigerators is the absence of moving parts at low temperature in PTs; this reduces vibrations which can spoil detector performance enhancing the lifetime of the cooler.

In the last 20 years, the so called Dry (or cryogen-free) Dilution Refrigerators (DDR) based on PT cryocoolers have started replacing wet dilution refrigerators and have become a standard. 
In these apparatus, the liquid He bath are usually replaced by a 2 stage PT providing cold stages at about 40 K and 4 K as shown in Figure~\ref{fig:pulsetube}. 
The 1K-pot is replaced by an extra heat exchanger located before the JT impedance; the returning warm $^3$He is cooled by the upgoing cold $^3$He vapour coming from the Still. 
This heat exchanger in combination with the JT expansion after the impedance is used to condense the gas. 
As the JT stage is less efficient than the 1K-pot, higher condensing pressures are needed, typically larger than 2\,bars during condensation and $\sim$ 0.5\,bars in operating mode. To reach these pressures an extra compressor is typically added on the condensing line and used during condensation.
This is also the solution used for the CUORE refrigerator.

\section{Development of large dilution refrigerators}\label{sec:LargeDRs}
%Several pioneering experiments in physics, such as precision observations of other rare events or dark matter, axion search, gravitational wave antennas, and quantum phenomena are in need of dilution refrigerators to reach the desired very low temperatures, which typically lie in the range of 10\,mK.
The next generation of experiments seeking to obtain precision observations of rare events (including neutrinoless double beta decay, dark matter, axions, gravitational waves, and quantum phenomena) are in need of dilution refrigerators to reach the desired temperatures, which typically lie in the range of 10-100\,mK.
The increasing size and mass of the parts to be maintained at very low temperature necessitates research towards high performance dilution refrigeration systems. 
 
In the past 40 years the cryogenic detector community developed DRs of increasing size capable of keeping tens of liters of experimental volumes at $\mathcal{O}$(10\,mK). 
During the same period very large cryostats were developed to host cryogenic resonant bars to search for gravitational waves.

\subsection{The mK ultra-cryogenics facilities before CUORE}\label{sec:ultra-cryogenics}

As briefly introduced in Section~\ref{sec:rare}, cryogenic rare event search experiments had to, since the very beginning, undertake the development of suitable cryogenic infrastructures for the calorimetric detectors. 
Several dilution units were installed in underground laboratories starting as early as late '80s.
A few of these facilities played a crucial role in the developments that followed. 

\begin{figure}[!t]
    \centering\includegraphics[width=0.7\textwidth]{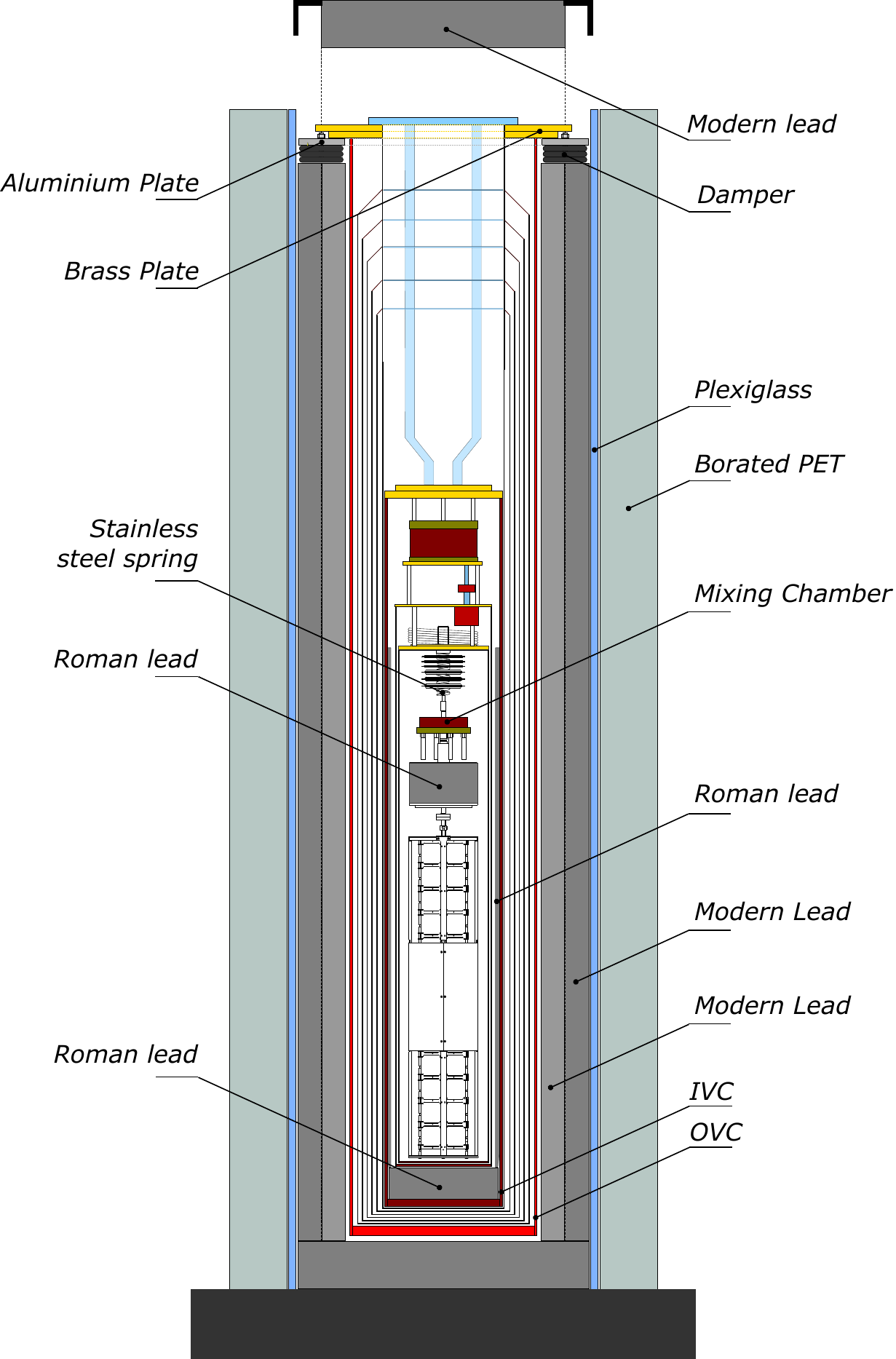}
    \caption{Sketch of the CUORE-0 cryostat and shielding~(not to scale). Modified based on a figure from~\cite{Q0-detector}.}
    \label{fig:Q0cryo}
\end{figure}

The MiDBD/CUORICINO/CUORE-0/CUPID-0 cryostat~\cite{Q0-detector}, installed at Laboratori Nazionali del Gran Sasso~(LNGS), was based on an Oxford 1000 DR~(capable of providing a cooling power of 1000 $\mu$W at 100 mK). 
The original experimental volume hosting the MiDBD-I experiment~\cite{MiDBD-IF} was about 7 liters, which was reduced to 5 liters during MiDBD-II to install an extra lead shield around the detector \cite{Arnaboldi_2002}. 
In 2001 an upgrade to the experimental setup allowed to fit the larger CUORICINO \cite{qino-2008} tower (42\,kg of TeO$_2$ crystals) in an experimental volume of about 23 liters, as can be seen in the sketch in Figure~\ref{fig:Q0cryo}. 
In 2002 CUORICINO was the first experiment to operate a large array of LTDs with a mass of the order of tens of kg of active absorbers. 
The same setup was than used for the CUORE prototype, CUORE-0~\cite{Q0_PRL}, and for the first CUPID demonstrator, CUPID-0~\cite{Azzolini_2018}

The CRESST experiment at LNGS was hosted in a cryostat based on an Oxford 1000 DR~\cite{angloher2002limits}. 
It's experimental setup was customized with a 1.5\,m long cold-finger which facilitated in enclosing the roughly 50 liter experimental volume with proper shielding.
This setup allowed collection of data up to 1 tonne/day of statistics over several years, even though in recent years the CRESST collaboration focused on very small ($\sim$24\,g) detectors to improve detector performance~\cite{Abdelhameed:2019hmk}.

The CDMS experiment initially developed a cryostat based on Oxford 400 DR which is capable of providing a cooling power of 400\,$\mu$W at 100\,mK. The cryostat was customized with a long horizontal cold finger to allow reverse~(top) opening shields for the detector installation. 
The original cryostat, installed in a shallow underground site (called SUF) at University of Stanford, had an experimental volume of $\sim$21 liters~\cite{CDMS-I}. 
The CDMS-II installation, in the Soudan mine, was based on an Oxford Kelvinox 400-S DR and followed the same reverse design of the original CDMS cryostat with major upgrades for background and noise reduction. 
The innermost shield had an experimental volume of $\sim$21 liters and hosted $\sim$4.6 kg of Ge and $\sim$1.2 kg of Si detectors~\cite{Agnese:2013rvf}.

The first EDELWEISS installation \cite{EDELWEISS2001}, located at the Laboratoire Souterrain de Modane~(LSM), was installed in a custom made DR with an experimental volume of $\sim$1.5 liters~\cite{EdelCryo96}. 
For the following phases, named EDELWEISS-II~\cite{EDELWEISS-II} and EDELWEISS-III~\cite{EDELWEISS-III}, a custom made DR with an horizontal cold finger was built with an experimental volume of about 50 liters. 
This allowed the easy installation of Ge detectors of masses up to 40\,kg. 

\subsection{The tonne-scale ultra-cryogenics facilities before CUORE}

Although, as discussed in Section~\ref{sec:ultra-cryogenics}, a large number of experiments were operated at ultra-low temperatures, until the construction of CUORE almost no other tonne-scale experimental apparatus had been cooled to such low temperatures. 
Only NAUTILUS and MINIGRAIL, two gravitational wave antennas operated in Frascati~(Italy) and Leiden~(Netherlands) respectively, had this kind of features, both of them hosted in cryostats equipped with dilution refrigerators.

\begin{figure}[!t]
    \centering
    \includegraphics[height=0.45\textwidth]{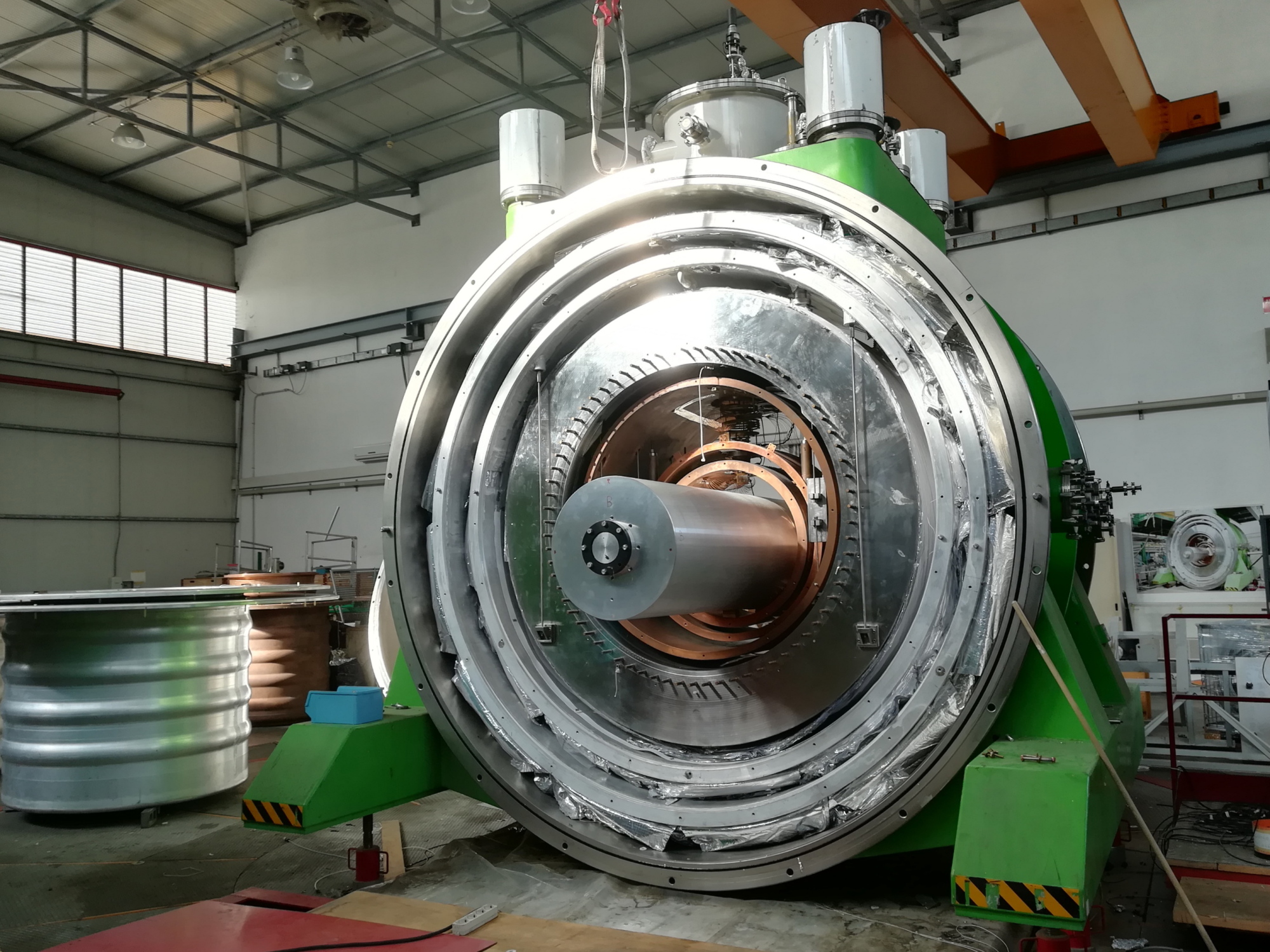}
     \includegraphics[height=0.45\textwidth]{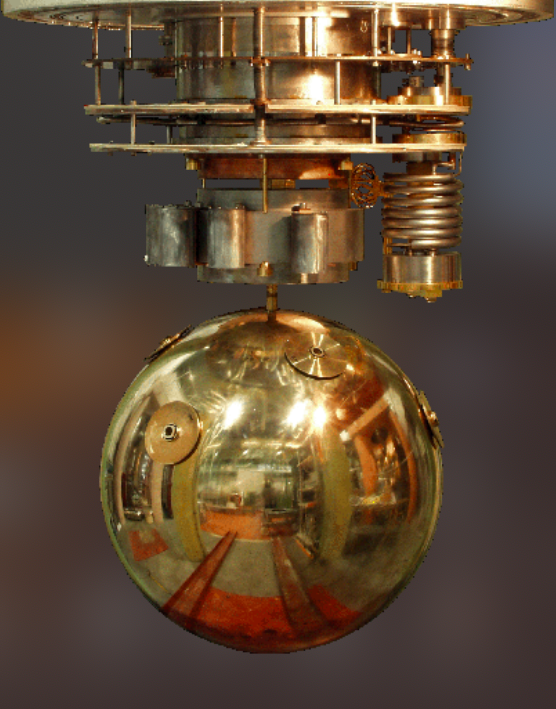}
    \caption{(Left) The disassembled NAUTILUS cryostat during its decommissioning~\cite{NautiL}. (Right) The MiniGRAIL spherical antenna and connected at a dilution refrigerator~\cite{MiniG}.
}
\label{fig:Nautilus_Minigrail}
\end{figure}

NAUTILUS~\cite{NAUTILUS} (Figure~\ref{fig:Nautilus_Minigrail}, left) was the first experimental apparatus at the tonne-scale to be cooled below 100\,mK. 
It consisted of  a cylindrical resonant bar made by an Aluminum alloy, 3.0\,m in length and 0.6\,m in diameter with a mass of 2300\,kg, refrigerated by a customized Oxford Instruments DR. 
It was part of a network that included the other resonant antennas EXPLORER~(at CERN) and AURIGA~(at INFN-LNL).
%It was designed to be operated at frequencies slightly higher than 900\,Hz. 
Although it was conceived to be cooled at sub-Kelvin temperature, after a first scientific run at 100 mK in 1998, technical motivations lead to continue the data taking keeping the bar at 2\,K, using pumped liquid helium.%~\cite{Nauti2000}.
The experimental runs of NAUTILUS ended in 2016.

MiniGRAIL (Figure~\ref{fig:Nautilus_Minigrail}, right) was the gravitational wave antenna developed by the Kamerlingh Onnes Laboratory group lead by G.~Frossati, in Leiden. 
Unlike the cylindrical bar of NAUTILUS, It consisted of a 1400\,kg CuAl alloy spherical detector, 68\,cm in diameter, and cooled at $T=\,65$\,mK by a DR~\cite{MINIGRAIL}. 
Compared to cylindrical bars, the spherical antennas would have the advantage of both detecting gravitational waves from any directions and having sensitivity to detect polarization. 
%MiniGRAIL operated at a frequency of 2.9\,kHz. 
The antenna was housed in a Kedel Engineering liquid helium cryostat, while the DR was developed by Frossati's group. 
MiniGRAIL was the coolest ton-scale experiment before CUORE. 
Despite this impressive result most of the science data of MiniGRAIL were obtained at 5\,K \cite{Gottardi2007}. 
As a part of the same network, the Mario Schenberg antenna~\cite{MARIOSCHENBERG} is a project funded by FAPESP~(Brazil) for a gravitational wave detector in Sao Paulo. 
Its design is based on the MiniGRAIL experimental apparatus and was commissioned down to the liquid helium temperature, but the antenna has not yet been cooled at sub-Kelvin temperatures~\cite{Aguiar2008}.

\section{The CUORE experiment}\label{sec:CUORE}
The Cryogenic Underground Observatory for Rare Events (CUORE) is an experiment that pioneered the use of a large array of cryogenic macro-calorimeters to search for neutrinoless double beta decay~\cite{CUOREREVIEW}.
The technique used for detection is quite simple: when a particle deposits energy in the absorber, the temperature increases.
By measuring the change in temperature one can, with appropriate calibration, infer the amount of energy deposited. 
The technique is quite powerful; with appropriate care one can achieve relative energy resolution better than  1\textperthousand\,over wide energy range (Table \ref{tab:reso}).

\begin{table}[!t]
    \centering
    \scalebox{1}{
    \begin{tabular}{l r r r}
    \hline 
    Experiment                     &  Energy      & Resolution FWHM  & Resolution \textperthousand \\
    \hline 
    LCLS-II~\cite{Morgan2019}      &  1.25\,keV   & 0.75\,eV  &  0.6 \\
    X-IFU Mission~\cite{Smith2012} &  5.9\,keV    & 1.6\,eV   &  0.3 \\
    CUORE R\&D~\cite{CANONICA_PHD} &  2.615\,MeV  & 2.8\,keV  &  1.1 \\
    CUORE R\&D~\cite{CCVR}         &  5.407~MeV   & 3.2\,keV  &  0.6 \\
    \hline
    \end{tabular}
    }
  \caption{\label{tab:reso} Energy resolutions obtained for low temperature detectors in the keV or in the MeV energy range.}
\end{table}

\begin{figure}[!t]
    \centering
    \includegraphics[width=\textwidth]{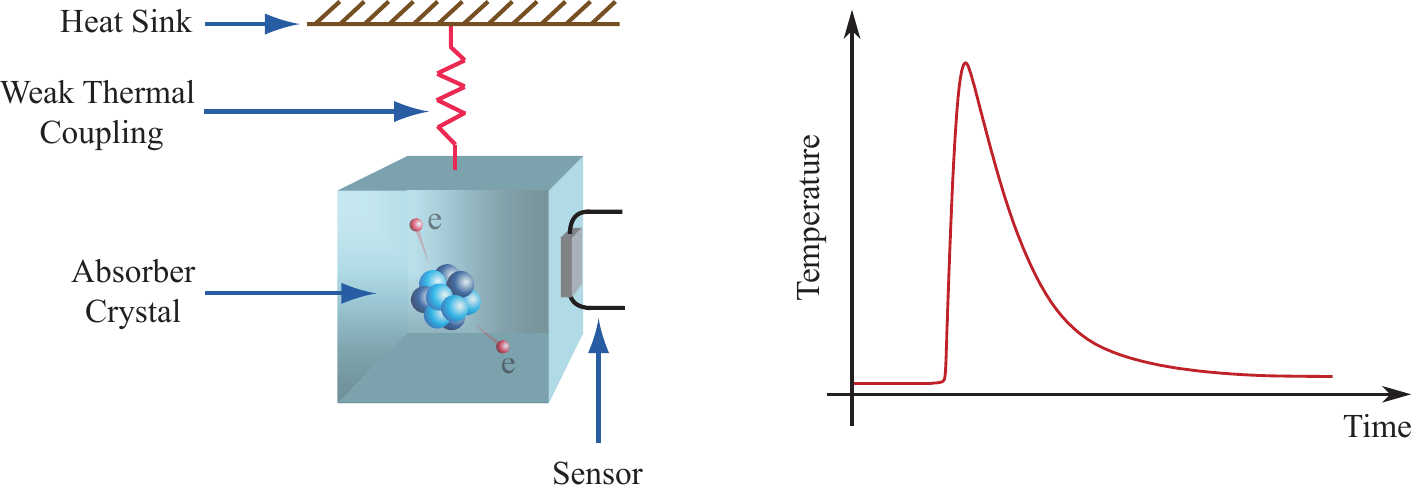}
    \caption{Conceptual implementation of a CUORE-style detector. Left: A decay occurring in the absorber causes the temperature to increase which is recorded by an attached thermometer. The absorber is coupled to a thermal bath which allows for gradual relaxation to base temperature. Right: Evolution of a thermal pulse in a CUORE-like detector.}
    \label{fig:bolometer-concept}
\end{figure}

Figure~\ref{fig:bolometer-concept} shows a sketch of the essential features of a CUORE-style detector, namely a massive absorber instrumented with a thermometer and thermally linked to a heat bath. 
A further remarkable property of this detection technique is that the absorber can incorporate the radioactive source being studied. This advantage is exploited in investigations of both $\beta$ decay~(as in the direct measurement of the neutrino mass~\cite{MARE_2005,HOLMES_2014}) and double beta ($\beta\beta$) decay. 
This approach has two main advantages: a simple design of the detector and a very high detection efficiency, since the decay under study occurs inside the detector.
In CUORE there is close to 750\,kg of TeO\textsubscript{2} absorber all of which has to be constantly maintained at a very low temperature of 10\,mK.

\subsection{Neutrinoless double beta decay}
\label{sec:ndbd}
% Neutrinos, despite being electrically neutral, feebly interacting, and \emph{almost} massless particles, continue to intrigue. 
%The discovery of neutrino flavour oscillations~\cite{PhysRevLett.81.1562} and subsequent precision oscillation experiments have demonstrated that neutrinos are massive, disproving a long-held assumption in the Standard Model of particle physics~\cite{PhysRevLett.89.011301,PhysRevLett.90.021802,PhysRevD.98.030001}.
%These tiny neutrino masses could be a tantalizing hint to the origin of the matter-antimatter asymmetry, one of modern physics' biggest mysteries.
% This asymmetry has the consequence that the matter which frames our existence remains in the universe today\,!  
%Leptogenesis via Majorana mass term has been hypothesized as a possible mechanism of the lepton-antilepton asymmetry in the early universe~\cite{FUKUGITA198645}.  
%For such an explanation to be viable, neutrinos must have mass, they must be Majorana particles, and lepton number must not be strictly conserved.  
%--- a quantum number attributed to fundamental particles --- 
%The discovery of non-zero nature of neutrino mass has intensified the search for neutrinoless double beta decay, which if observed would simultaneously show that neutrinos are Majorana particles and that lepton number is violated. 
%The implications as a smoking gun for leptogenesis make these searches particularly compelling.
Neutrinoless double beta decay~\cite{Furry_1939} is a hypothesized lepton-number-violating rare nuclear process not predicted by the Standard Model. 
This decay may be considered a neighboring reaction to a Standard Model-allowed process called two neutrino double beta~($\twonu$) decay in which two neutrons in a nucleus simultaneously decay into two protons, two electrons and two electron antineutrinos: $2n\longrightarrow 2p + 2e^{-} + 2\bar{\nu}_{e}$.  
Being a second-order weak interaction, \twonu~decay is exceedingly rare, exhibiting half-lives in the range of $10^{18}-10^{21}$ years. 
This lepton number conserving reaction has been observed in some even-even nuclei such as $^{76}$Ge, $^{100}$Mo and $^{136}$Xe~\cite{saakyan2013two} .
%The concept of lepton number conservation is illustrated in this reaction. 
%We assign a lepton number of $+1$ to the electrons and $-1$ to the antineutrinos, thus the change in lepton number is zero for this process. 
In the neutrinoless mode, two neutrons decay into two protons and two electrons with no neutrinos emitted: $2n\longrightarrow 2p + 2e^{-}$. 
% Lepton number increases by +2 in this process. 
Theoretically this process can proceed, for example, by exchange of a virtual light~\emph{Majorana neutrino}.  
%Two neutrons decay, producing two protons, and two W-bosons. 
%One W-boson decays into an electron and a~\emph{Majorana neutrino}, a state which is mostly thought of as the antiparticle , i.e., the right-handed state but which also contains a small admixture of what we think of as the particle, i.e., the left-handed state.  
%The size of this admixture depends on the mass of the neutrino.  
%The left-handed component can be absorbed by the second W-boson causing it to emit an electron. 
%Effectively, two neutrons decay without emission of any neutrinos in the final state.  

The experimental signature of \zeronu~decay is expected to be relatively robust. 
Since almost all the energy in the decay will be carried by the final state electrons, measuring the sum energy of these electrons would produce a spectrum which is simply a peak at the Q-value of the decay. 
On the other hand,  measuring the sum energy of the electrons emitted in \twonu~decay produces a continuous spectrum distributed from zero-energy up to the Q-value as some energy will be carried away by the undetectable antineutrinos.  
Thus the search for \zeronu~decay boils down to a search for an anomalous peak at the Q-value of the transition in the spectrum of decays produced by a sample which is unstable to double beta decay. The experimental sensitivity ($S^{0\nu}$), defined as the decay time corresponding to the minimum number of detectable events above background ($B$), given by

\begin{equation}\label{eq:sens} 
%S^{0\nu}\propto\epsilon\cdot\mathrm{i.a}\cdot\sqrt{\frac{M\cdot T}{B \cdot \Gamma}}
S^{0\nu}=\ln{2}\cdot\epsilon\cdot\frac{\mathrm{i.a.}}{A}\cdot\sqrt{\frac{M\cdot T}{B \cdot \Gamma}}
\end{equation}
where $\mathrm{i.a.}$ is the isotopic abundance of the a chosen $\beta\beta$-emitter isotope, $M$ total sensitive mass, $A$ its molecular mass, $\epsilon$ the efficiency of the detector, $\Gamma$ the energy resolution (around the $Q$-value), $B$ is the background index, expressed in counts/(keV$\cdot$kg$\cdot$y), and finally, $T$ the measurement time. The product of $M$ and $T$ is defined as \emph{exposure} of the detector. To obtain the best sensitivity a double beta decay experiment must have a very large mass, high efficiency, a good resolution, a long measurement time, a very low background, and the chosen $\beta\beta$-emitter isotope should have a high natural isotopic abundance. Low temperature detectors, combined with a large cryostat able to provide high cooling power, low noise environment, and extremely low radioactivity content, represent an experimental approach able to fulfill all of these requirements at once.   

For a review on the current status of the search for neutrinoless double beta decay we refer the reader to the review article~\cite{NDBD_Review}.

\subsection{From single crystal to CUORE}

The development of the CUORE detector has proceeded in multiple phases. 
The long chain of cryogenic experiments based on TeO\textsubscript{2} crystals started almost 30\, years ago~\cite{CUORE_PRECURSOR}.  Over the years several detector setups were developed and hosted in the cryogenics facility described in Section~\ref{sec:ultra-cryogenics}, the last one being CUPID-0~\cite{Azzolini_2018}, the demonstrator array for the CUPID (CUORE Upgrade with Particle Identification) experiment~\cite{CUPID_preCDR_2019}. 

The first set of measurements with TeO\textsubscript{2} cryogenics detectors was carried out with single crystal detectors with gradually increasing sizes: from 5.7 to 334\,g~\cite{Alessandrello_1992_6g,GIULIANI_1991,Alessandrello_1992,Alessandrello_1993,Alessandrello_1994}. The first array of TeO\textsubscript{2} detectors, made using four $3 \times 3 \times 6$\,cm\textsuperscript{3} crystals, each with an average weight of 340\,g~\cite{Alessandrello_1995} was constructed in 1994. This was the first of its kind cryogenic detector with a combined mass exceeding 1\,kg. In 1997, a tower made of 20 detectors (5 floors of 4 crystals each) was assembled and cooled at LNGS.  Each individual absorber was a $3 \times 3 \times 6$\,cm\textsuperscript{3} crystal with a total active mass of about 6.8\,kg. The experiment --- which was later named MiDBD~(Milan Double Beta Decay)--- was the largest operating cryogenic detector of its time. The CUORICINO experiment~\cite{qino-2008} was the first step towards realizing a large array calorimeter experiment. The detector consisted of 44 large crystals ($5 \times 5 \times 5$\,cm\textsuperscript{3}) and 18 smaller crystals ($3 \times 3 \times 6$\,cm\textsuperscript{3}) that were borrowed from the MiDBD array. Crystals were arranged in 13 floors, with 11 floors housing 4 large-size crystals each and the remaining two floors housing 9 smaller ones each. The total mass of TeO\textsubscript{2} was 40.7\,kg, corresponding to about 11\,kg of \textsuperscript{130}Te~\cite{qino-2008}. CUORICINO proved the feasibility of scaling to a 1000-crystals low-temperature array, achieving a good energy resolution and efficiency. CUORE-0~\cite{Q0-detector} was a single CUORE-like tower with the goals to validate the cleaning and assembly procedures designed for CUORE, and to perform a high statistics assessment of the improvements in the background mitigation. Composed of 52 natural TeO\textsubscript{2} $5 \times 5 \times 5$\,cm\textsuperscript{3} cubic crystals, the CUORE-0 detector was divided into 13 floors of 4 crystals each. The detector had 39\,kg of active mass, corresponding to about 10.8\,kg of \textsuperscript{130}Te. The tower was constructed following the same procedures and using the materials selected for CUORE. The detector installation was completed in late 2012 and collected data from March 2013 until March 2015. Table \ref{tab:exp_list} summarizes the main features of the  TeO\textsubscript{2} arrays  over the years with the results in terms of energy resolution at the $\beta\beta$ Q-value, measured background in the region of interest (ROI), limits on the $0\nu\beta\beta$ half-life, and neutrino mass sensitivity.   

\begin{table}[!t]
    \scalebox{1}{
    \begin{tabular}{l r c r r r r }
      \hline
      Experiment	& \multicolumn{3}{c}{Running period}	
                    & Crystals	
                    & Mass	
                    & Exposure\\
                    & \multicolumn{3}{r}{}
                    &
                    & [kg]	
                    & [kg\,yr]\\	
      \hline
      MiDBD~\cite{Arnaboldi_2002} & Apr 1998 & - & Dec 2001		
                                  & 20	
                                  & 6.8		
                                  & 4.25\\
      
      CUORICINO~\cite{qino-2011}  & Mar 2003& - & Jun 2008		
                                  & 62	
                                  & 40.7		
                                  & 71.4\\
      
      CUORE-0~\cite{Alfonso:2015wka}& Mar 2013& - & Mar 2015		
                                    & 52	
                                    & 39.0		
                                    & 35.2\\
      
      CUORE~\cite{CUORE_NATURE_2021}& May 2017& - & Dec 2020		
                                    & 988	
                                    & 742 
                                    & 1038.4\\
      \hline
      &&&&& & \\
      &&&&& & \\
      \hline
      Experiment	& FWHM\,@\,$Q_{\beta\beta}$
                    & \multicolumn{2}{r}{bkg in the ROI}	
                    & \multicolumn{2}{r}{$t_{1/2}^{\,0\nu}$ {\scriptsize (90\%\,C.\,L.)}}	
                    &$m_{\beta\beta}$	\\
    
      	            & [keV]
      	            &\multicolumn{2}{r}{[c/keV/kg/yr]} 
      	            &\multicolumn{2}{r}{[yr]}		    
      	            &[ev] \\	
      	
      \hline
      
      MiDBD~\cite{Arnaboldi_2002} & $5-15$
                                    &\multicolumn{2}{r}{0.3}
                                    &\multicolumn{2}{r}{$2.1 \cdot 10^{23}$}
                                    &$1.6$\\ 
                                    
      CUORICINO~\cite{qino-2011}    & $5.8 \pm 2.1$
                                    &\multicolumn{2}{r}{0.15}
                                    &\multicolumn{2}{r}{$2.8 \cdot 10^{24}$}
                                    &$0.43$\\
      CUORE-0~\cite{Alfonso:2015wka} 
                                    &$4.9 \pm 2.9$
                                    &\multicolumn{2}{r}{$0.058 \pm 0.004$}
                                    &\multicolumn{2}{r}{$2.8 \cdot 10^{24}$}
                                    &$0.36$\\	
      CUORE~\cite{CUORE_NATURE_2021}&$7.8 \pm 0.5$
                                    &\multicolumn{2}{r}{$0.0149 \pm 0.0004$}
                                    &\multicolumn{2}{r}{$2.2 \cdot 10^{25}$}
                                    &0.186\\
    \hline
    \end{tabular}
    }
  \caption{\label{tab:exp_list} List of the cryogenic TeO$_2$ experiments searching for the of $^{130}$Te with the main characteristics and obtained results reported. Values for $t_{1/2}^{\,0\nu}$ are lower limits.}
\end{table}

The need to improve the performances of thermal detectors has led, over the years, to several updates to the cryogenics facility and to the detector design. These updates represented the starting point of the design of the CUORE experiment and its cryostat. On the cryostat side, the noise induced by vibrations represents a serious problem for large thermal detectors. This noise affects the energy resolution of the detectors reducing the experimental sensitivity. This vibrational noise has been studied with dedicated measurement~\cite{Pirro_2000} and then mitigated by the design and construction of an optimized two-stage damping system~\cite{Gorla_2004,Pirro_2006}, used to decouple the detector tower to the cryostat plate. During the MiDBD operation, in order to reduce the background contribution due to the radioactive contaminations of cryostat and  
to shield the detector from  the intrinsic radioactive contamination of the dilution unit materials, a new framed Roman lead
shield was placed around the tower, inside the cryostat itself. Roman lead of archaeological origin  has unique properties in terms of low intrinsic radioactivity, given its complete depletion of the contaminant $^{210}$Pb~\cite{Pattavina:2019pxw}, which makes it ideal for a rare event experiments. Roman lead shield around and below the detector, and modern low-radioactivity lead shield above the detector, were also designed and installed in the CUORICINO, CUORE-0 and CUORE experiments. The total mass of the roman lead kept at cryogenic temperature increased from few tens of kg (MiDBD, CUORICINO, and CUORE-0) to 4.5\,tonnes (CUORE). To host this amount of lead, together with the detector array, a large cryostat with a proper cooling power and with very low levels of intrinsic radioactivity becomes a fundamental requirement.

On the detectors side, performance is limited by the non-uniformity of the detector response and by the background rate in the ROI, attributed primarily to two main sources: radioactive contamination of the cryostat or its gamma shields, and degraded alpha particles from the surfaces of the crystal holder~(most likely from copper). To rule out these problems, in parallel with the CUORICINO data taking, several investigations were carried out in view of the forthcoming CUORE. To reduce the contribution to degraded $\alpha$s, from either the crystal surfaces or the support structure parts~(copper and PTFE), a new design for the detector structure was proposed. The new solution reduced the amount of copper and the copper surfaces facing the TeO\textsubscript{2} crystals by a factor $\sim 2$. New production and certification protocols were developed to reduce the bulk and surface contaminations in the CUORE crystals~\cite{JCG312}. To mitigate the background surface contamination of inert detector materials new cleaning procedures were developed and tested~\cite{TTT}. The selected copper cleaning procedure consisted of 4 steps: tumbling, electro-polishing, chemical etching and magnetron plasma cleaning. 
In CUORICINO the chip-to-crystal gluing was a manual, labor-intensive procedure that caused wide variability in behaviour among all the detectors. To reduce this problem a semi-automated system was developed to achieve far more precise and reproducible chip-absorber interfaces. This system included a six-axis articulated robotic arm to lift and position the crystals, and a three-axis Cartesian robot to dispense glue dots on the semiconductor chips via a pneumatic dispenser~\cite{RUSCONI_PHD}. The CUORE Tower Assembly Line~(CTAL) allowed to assemble the ultra-clean detector parts into the ultra-clean CUORE towers avoiding possible recontamination~\cite{CTAL}.

The CUORE-0 results showed that the new assembly line and procedures were able to provide robust and reproducible detector characteristics leading to excellent detector performance with an effective background suppression. In fact, compared with CUORICINO, the background rate of CUORE-0 was a factor of three lower in the ROI and factor of seven lower in the $\alpha$-region (Table \ref{tab:exp_list}). Gamma contaminations from within the cryostat materials contributed to the remaining excess, and formed an irreducible background. These results demonstrated that the procedures applied to CUORE-0 matched the goals of CUORE.

%CUORE-0~\cite{Q0-detector} was a single CUORE-like tower with the goals to the cleaning and assembly procedures designed for CUORE, and to perform a high statistics assessment of the improvements in the background mitigation. Composed of 52 natural TeO\textsubscript{2} $5 \times 5 \times 5$\,cm\textsuperscript{3} cubic crystals, the CUORE-0 detector was divided into 13 floors each composing of 4 crystals. The detector had 39\,kg of active mass, corresponding to about 10.8\,kg of \textsuperscript{130}Te.  The tower was constructed following the same procedures and using the materials selected for CUORE. The detector installation was completed in late 2012 and collected data from March 2013 until March 2015. After an initial optimization phase~\cite{Q0-InitialPerformances}, CUORE-0 achieved an effective mean FWHM energy resolution of ($4.9 \pm 2.9$)\,keV near the 2615\,keV \textsuperscript{208}Tl line with the final \textsuperscript{130}Te exposure of 9.8\,kg$\cdot$yr~\cite{Q0_PRL}. 

\subsection{The CUORE detector array}
\label{sec:det_array}
CUORE is a ton-scale cryogenic detector that consists of an array of 988 natural TeO\textsubscript{2} cubic crystals. 
Each cube has a 5 cm dimension and weighs $\sim$750\,g resulting in a total mass of 742\,kg of TeO\textsubscript{2} to be cooled to mK temperatures. 
With this mass, the CUORE array is almost twenty-fold larger than CUORE-0.  
%The crystals are assembled into 19 towers; each tower containing 52 crystals arranged into four-crystal floors and stacked 13 floors high in a copper support frame. 
The crystals are assembled into 19 towers following the CUORE-0 design.
Each crystal is instrumented with a neutron transmutation doped~(NTD) germanium thermistor~\cite{Haller1984} to record thermal pulses, and a silicon heater~\cite{Alessandrello1998,Andreotti2012} that provides reference pulses for thermal gain stabilization. 
The crystals are held in the copper frames by polytetraflourethylene (PTFE) spacers which provide a weak thermal link to the heat bath. 
The towers hang as a close-packed array from a tower support plate (TSP) thermally linked to the 10\,mK stage of CUORE cryogenic system.

The installation of all towers in the cryostat was completed in August 2016, followed by the cooldown to the base temperature at the beginning of 2017. 
%After an initial optimization phase~\cite{CUORE_PRL_2018} a total exposure of 372.5\,kg$\cdot$yr~\cite{CUORE_PRL_2020} of TeO\textsubscript{2} were accumulated during a period between May 2017 and July 2019.
After an initial optimization phase~\cite{CUORE_PRL_2018} a total exposure of 
1\,tonne-year worth of TeO\textsubscript{2} data was accumulated during a period between May 2017 and December 2020.
%The effective mean FWHM energy resolution at the 2615\,keV \textsuperscript{208}Tl line is 7.73(3)\,keV. 
%\ptsnote{need to refer to the new 1 ton paper if/when it comes out.}
The effective mean FWHM energy resolution at the 2615\,keV \textsuperscript{208}Tl line is $(7.78\pm 0.03)$\,keV.
%The background of rate 0.0138~(7)\,counts/keV/kg/yr in the ROI corresponds to an improvement by a factor 5 with respect to CUORE-0.
The background rate of $(1.49\pm0.04)\cdot 10^{-2}$~counts/keV/kg/y in the ROI corresponds to an improvement by a factor $\sim$5 with respect to CUORE-0. The stable conditions provided by the cryogenics facility allowed for continued data taking with minimal human intervention and this were highly beneficial during the COVID-19 lock-down.
% This enabled CUORE to reach a total exposure of 1\,tonne$\cdot$year of TeO\textsubscript{2} in December 2020. 
While \zeronu~remains the focus of the CUORE experiment, the large target mass and the ultra-low backgrounds make CUORE an excellent detector to search for exotic and rare decays~\cite{CUORE_RARE}, as well as to study the interactions of WIMPs and solar axions in the detectors described later in section~\ref{sec:future}.
%the why and the how of this new observatory at the frontier of cryogenic particle detection.

% \begin{comment}
% \newpage
\section{The CUORE cryostat} \label{sec:cuore_cryostat}

As discussed in the previous sections, one of the main challenges of the CUORE project has been to realize a large cryogenic infrastructure capable of maintaining the  full CUORE detector at a stable temperature of $\sim$ 10\,mK on the timescale of 5--10 years. The fact that the CUORE cryostat has been designed to host an experiment aiming to be highly sensitive to \zeronu~\cite{Alduino:2017pni} made the design and realization of such an infrastructure even more challenging, as one must also account for the reduced background and low noise requirements. 
% This unusual feat has been achieved for the first time at CUORE.
This section provides an overview of the driving motivations and of the solutions adopted in the design and development of the CUORE cryostat, while Section~\ref{sec:commissioning} goes into the details of the challenges faced in realizing the facility and the technical strategy developed during the cryostat commissioning.

\subsection{Design and realization}

The design and realization of the CUORE cryostat has been one of the largest challenges in the history of experimental cryogenics, having to satisfy a number of very stringent requirements. 
In this section, the guidelines for CUORE cryostat development are specified separately for clarity, but as it will be evident, they are the result of a complex evaluation of different contributions and of choices that were made while prioritizing the final scientific goal of CUORE. The main requirements for the experimental sensitivity are described in Section~\ref{sec:ndbd} and have direct impact on the detector development and on the cryostat design: namely, the need for a large mass (large experimental volume), long live-time (dry and stable cryostat), good energy resolution (low mechanical noise), and low backgrounds (material selection and shielding).
Based on these requirements, the design and development of the CUORE cryostat followed three main guidelines:
\begin{enumerate}
\item Cryogenic Performance: guarantee cooling power to cool-down the experimental volume to a temperature of $\sim$10\,mK, and maintain stability to within 0.1\,mK on a multiyear timeline;
\item Low backgrounds: provide a low radioactive-background environment through proper selection of the location, materials, set of radiation shields;
\item Mechanical noise suppression: keep the experimental volume isolated from mechanical vibrations, of human or natural origin, to allow optimal detector performance.
\end{enumerate}
These guidelines were very often in conflict with each other and the cryostat design choices were a balance between the competing needs. 

% Although in the following text these guidelines are described separately for clarity, as it will be evident, they are the result of a complex evaluation of different contributions and of the choices that were made prioritizing the final scientific goal of CUORE. The main requirements for the experimental sensitivity are described in Section~\ref{sec:ndbd} and have direct impact on the detector development and on the cryostat design: the need for a large mass (requires large experimental volume), long live-time (requires a dry and stable cryostat), good energy resolution (requires low mechanical noise), and low backgrounds (places constraints on material selection and shielding), are widely described below.  
% The main demands for the cryostat were:\\
%-	to cool a complex of 988 crystals with a total volume of the order of a cubic meter and a total weight at the ton scale, at a fixed temperature around 10\,mK, stable within 0.1\,mK;\\
%-   to carry out the detector signals and control the heaters via 2600 wires without heating the crystals;\\
%-	to keep the detector isolated from every mechanical noise coming both from the cryostat subsystems and from the ambient (including earthquakes) as much as possible;\\
%-	to keep the detector shielded from any source of electromagnetic noise;\\
%-	to keep the detector shielded from radioactive background, both produced from the ambient and from the cryostat itself;\\
%-	to keep it all working for several years of data taking.\\
%-	to be compliant with the earthquake safety requirements for the region;\\

\subsubsection{Cryogenic performance}
The expected performance of the CUORE cryostat, namely to cool the 988 TeO$_2$ crystals to a temperature of $\sim$ 10 mK with a stability of $\pm$~0.1 mK over years of data taking, is contingent on it satisfying the following requirements.
The main challenge is the realization of the experimental volume needed to host such a large number of detectors. The cryostat is composed of nested vessels (discussed in section~\ref{sec:cryostat}) and the volume of the innermost temperature shield, the so called 10 mK vessel, is $\sim$1 m$^3$ and hosts installation of the complete detector array and the Top Lead shield designed to protect the detectors from radiation generated in the upper part of the cryostat or above the cryogenic facility.
The second crucial thermal load is contributed by the detector wiring system which is composed of the read-out of NTD thermistors, auxiliary thermometers, and other channels in the experimental volume from inside the cryostat to the front-end electronics, and the wires used to bring the heat pulses to the detector heaters.    
The NTD thermistor signals from each of the 988 detectors are read through a twisted pair of NbTi wires~\cite{Andreotti_2009,Giachero_2013} that run across the cryostat from the 10\,mK plate to room temperature plate and are extracted through so called Fisher boxes, from which the signal is brought to the front-end electronics. In addition to these 1976 wires, extra set of wires had to be supplemented to bring the heating pulse to the detector heaters, and to read out auxiliary thermometers and channels in the experimental volume. 
% A total of 2600 NbTi wires, arranged in twisted pairs, were installed in the CUORE cryostat.
These pairs run from the 300\,K plate to the 10\,mK plate and are thermalized at each temperature stage. On the 10\,mK plate, specially designed connector boxes allow to connect the readout chain to the signal Cu-pen strips~\cite{Brofferio_2013} coming from the detectors. 
The third relevant load is from the internal lead shields, designed to protect the detector from the cryostat contamination as described in the next section. 

\subsubsection{Low backgrounds}
As described in Section~\ref{sec:ndbd}, reducing backgrounds is crucial in enabling CUORE to reach its scientific goal. The CUORE background budget \cite{Alduino2017epjc} demands a total background rate, in the ROI for 0$\nu \beta\beta$ of $^{130}$Te, of the order of 0.01 counts/keV/kg/yr. The major contributions to the background are:
\begin{itemize}
    \item Natural radioactivity outside the cryostat, which is suppressed by means of the external shielding
    \item Natural radioactivity and contamination in the detector's crystals and structure, which was minimized during construction by the use of stringent material selection and screening processes for the detector's components
    \item Natural radioactivity and contamination in the cryostat structure
\end{itemize}
The last source of radioactivity was mitigated both by a stringent material selection campaign and by the use of internal shields to protect the detector. All the cryostat plates and vessels have been made out of copper from a specially selected batch with low radioactive content. Additionally, the dilution unit and the pulse tubes are commercially produced and made using materials that cannot be easily modified~(e.g., sintered silver was used for the DU mixing chamber). 
%To prevent the radiation produced by possible external contaminants from reaching the detector, lead shields have been inserted inside the cryostat around the experimental volume. 
% A 6 cm thick lateral lead shield, made out of ancient ``Roman" lead \cite{AlessandrelloRoman} of archaeological origin is installed between 4K and 600 mK. The Top Lead shield\ptsnote{Paragraph is unfinished.} is sandwiched between two plates copper and is composed of five 6 cm disks of pure commercial lead. 

\subsubsection{Mechanical noise suppression}
Isolation and decoupling of the detectors from all possible sources of mechanical noise is another important requirement of the CUORE cryostat. As described in Section~\ref{sec:CryoDet}, one of the main components of the resolution degradation in low temperature detector is the so called thermal noise. This noise is generated by mechanical vibrations that convert into heat directly in the absorber through friction. Another relevant contribution to noise is microphonics, which is also induced by mechanical vibrations on different parts of the detector or of the readout chain. 
The design goal of CUORE was to obtain a resolution of the order of 5\,keV (FWHM) in the ROI. 
Despite the difficulty in directly correlating mechanical vibrations with the detector baseline RMS broadening, a strong effort has been put in place to keep this noise well below 5\,keV.
To minimize these effects in the CUORE cryostat, various solutions were adopted at each thermal stage. More details are described in \cite{Alduino:2019xia}. The detector --- being most critical part of the experiment --- is decoupled from the cryostat, hanging from a mechanically insulated structure (called Y-beam) and is thermalized to the cold stages by means of very soft mechanical connections. Unfortunately some sources of mechanical noise are strictly connected with the cryostat, such as the PTs. For this reason, an active noise cancellation~\cite{ActiveNoise} system was developed and implemented in order to minimize noise contributions from vibrations induced by the unavoidable gas expansion inside the PTs. 
\ptsnote{Seems like an abrupt ending. May need a bit of expansion.}
%However, this cryostat being a one-of-a-kind prototype, during the installation and the commissioning phase, a number of details have been modified with respect to the initial design, as a result of a continuous improvement and optimization.

\subsection{The structure of the CUORE cryostat}
\label{sec:cryostat}
\begin{figure}
    \centering
    \includegraphics[height=0.8\textwidth]{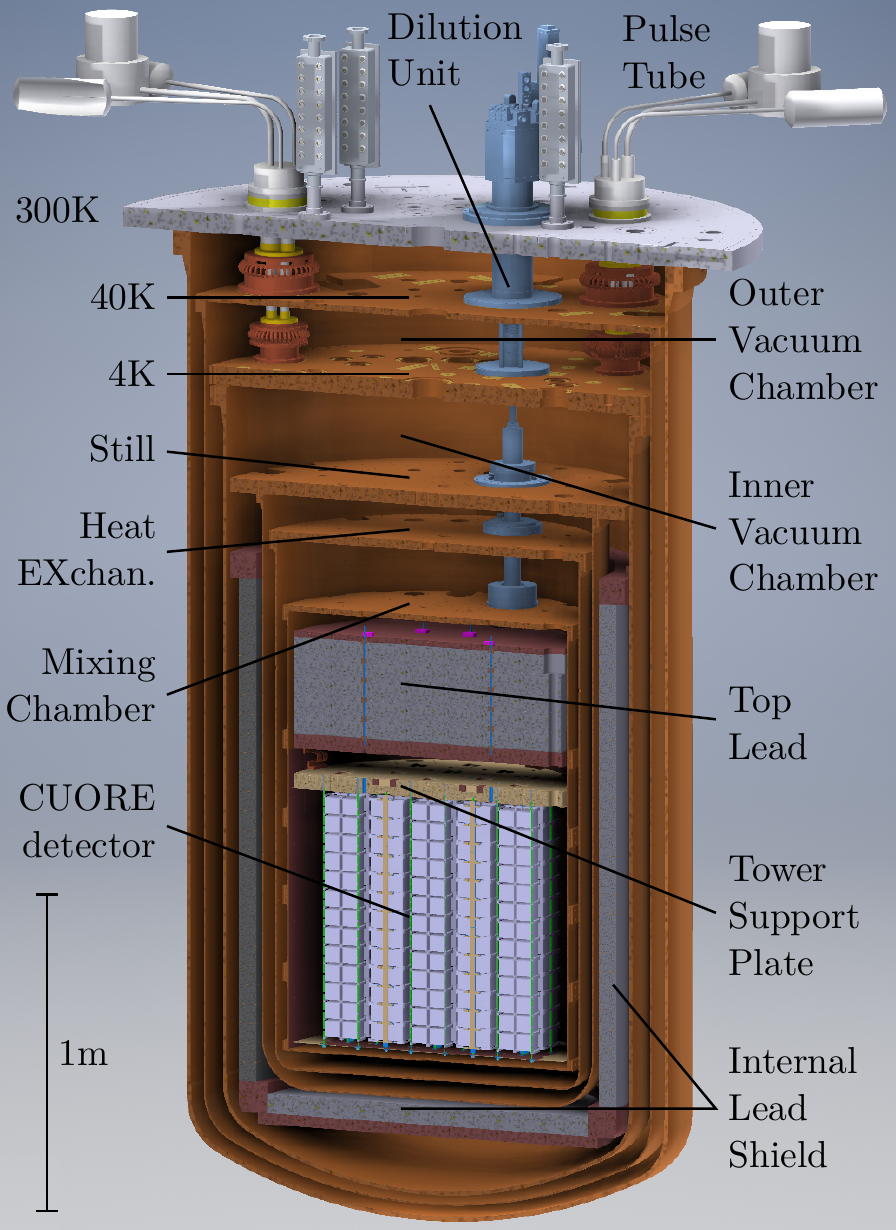}%Cryostat.png}
    \caption{Rendering of the CUORE cryostat. The cryostat is composed of six nested vessels and each stage is characterized by the approximate temperature stages (300\,K, 40\,K, 4\,K, 600\,mK, 50\,mK, 10\,mK), characterized by 5 PTs and by the DU.}
\label{fig:cryostat}
\end{figure}

The CUORE cryostat~(Fig.~\ref{fig:cryostat}) was designed following the above guidelines. Being a groundbreaking apparatus with very strict custom requirements, the design, manufacturing supervision, assembly, and commissioning of the CUORE cryostat have been carried out directly by the CUORE cryogenics group.

The cryostat is composed of six nested vessels as shown in the Fig.~\ref{fig:cryostat}.
The upper plate --- whose thickness (62\,mm) was calculated by mean a finite element analysis to support the weight of all the cryostat vessels --- is made of stainless steel (SS304L). To minimize the radioactive contamination, all other plates and vessels down to the 50\,mK stage are made of oxygen-free electrolytic copper (OFE Cu) at 99.99\% purity grade, with low H$_{2}$ content. For the bigger parts working at 10\,mK (MC plate and vessel, TSP, and detector frames), an ETP1 copper alloy named NOSV copper~\cite{nosv} has been chosen. This material exhibits very high conductivity at low temperatures (residual-resistance ratio $RRR>$400), has more stringent bulk radioactivity limits than OFE Cu, and low H$_{2}$ content. H$_{2}$ contamination becomes important at mK temperatures because of the heat released from the spontaneous conversion of ortho-hydrogen to para-hydrogen at these temperatures~\cite{Schwark1983}.

The 300\,K and the 4\,K vessels are vacuum-tight, enclosing the OVC~(5.86\,m$^{3}$) and the IVC (3.44\,m$^{3}$), respectively. A partition of the cryostat volume into two vacuum spaces is necessary in order to allow pre-cooling of the detector and filling of the IVC with exchange helium gas while keeping the OVC under vacuum. The IVC contains the experimental space, the lateral and the Top Lead shields as shown in Fig~.\ref{fig:cryostat}. A buckling analysis was carried out to determine the thickness of OVC and IVC vessels, also taking into account the differential pressure, positive and negative, that the two volumes can withstand during the gas filling. The OVC is sealed with elastomer o-rings, while the IVC/OVC interfaces are sealed with indium at the small interfaces and a tubular metallic seal~\cite{helioflex} at the 4\,K plate/vessel interface. TiAlV washers were mounted on the bolts for the indium seals, to compensate for the differential thermal contractions between copper of the flanges and the stainless steel of the bolts.

\begin{figure}
    \centering
    \includegraphics[height=0.5\textwidth]{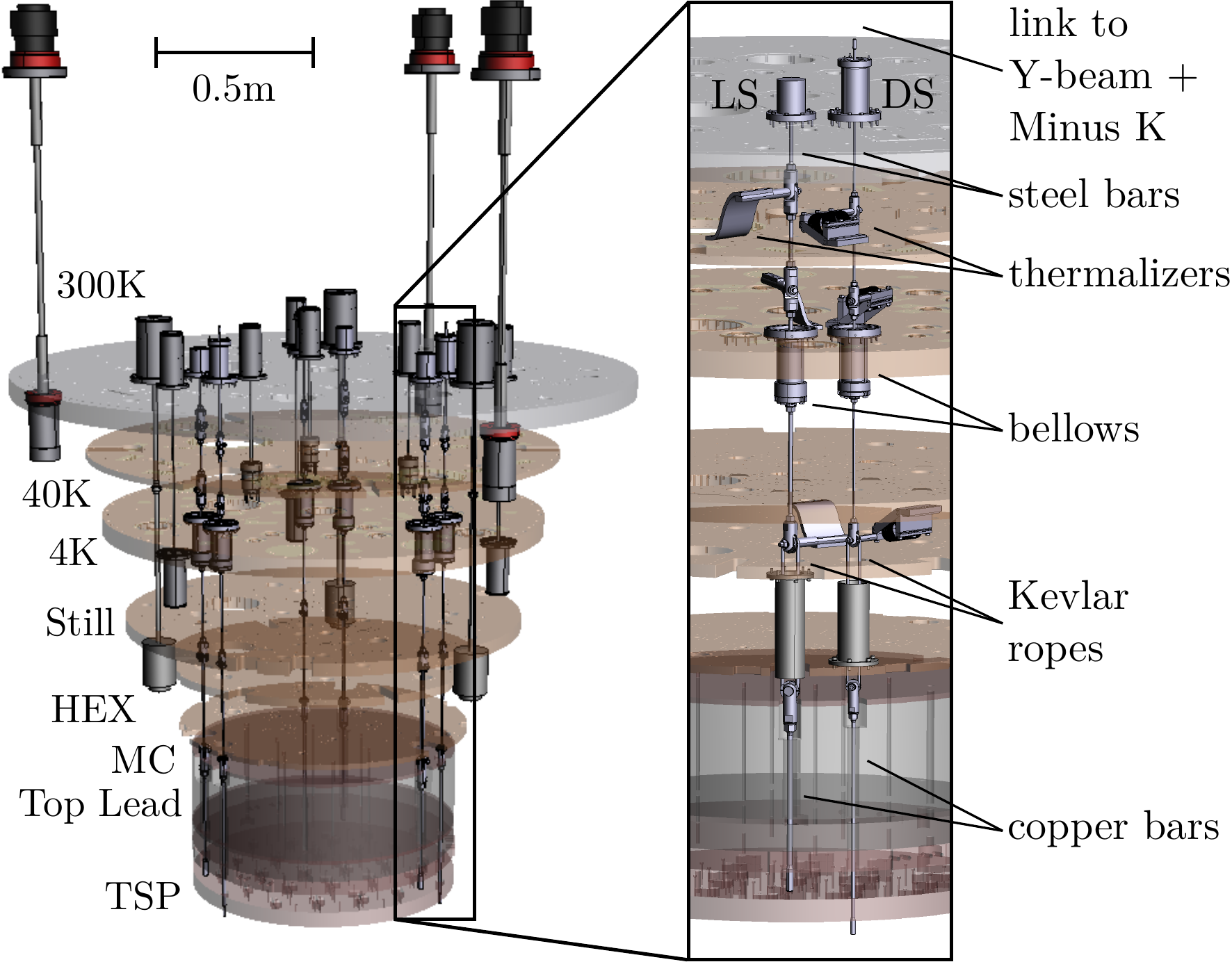}%Bars.png}
    \caption{Render of the supporting structure of the inner vessels.
    The MSP (at the top, not shown in the figure) supports the 300\,K vessel from above by means of three steel ropes. The 300\,K plate in turn supports the 40\,K, 4\,K and the Still plates, independently. The HEX plate is sustained by the Still plate and supports the MC (10\,mK) plate.
    }
\label{fig:bars}
\end{figure}

To absorb the radiation emission from the cryostat plates and from the upper external environment a 900(d)x300(h)\,mm, 2100\,kg lead plate named Top Lead has been inserted above the detector. It is placed below the MC plate, just above the TSP, which is the copper disk that mechanically supports the detector.
Laterally and below the detector, 60\,mm thick (4540\,kg in mass) ancient Roman lead (named Internal Lead Shield, ILS) is placed in between the 4\,K and the Still vessels and cooled at 4\,K. ILS is in effect the lead closest to the detector, so it is extremely important to choose a material with very low radioactivity.
A detailed view of the supporting structure of the inner vessels is depicted in Fig.~\ref{fig:bars}.
%This lead was found in wrecks of Roman ships sunk in the Mediterranean Sea, and exhibits extremely low levels of $^{210}$Pb activity~\cite{AlessandrelloRoman}, having been left free to decay for two thousand years.\ptsnote{Some repitition from `From single crystal to CUORE' section.}

% A detailed view of the supporting structure of the inner vessels is depicted in Fig.~\ref{fig:bars}. The Main Support Plate (MSP, at the top, not shown in the figure) supports the 300\,K vessel from above by means of three steel ropes. The 300\,K plate in turn supports the 40\,K, 4\,K and the Still plates, independently. The HEX plate is sustained by the Still plate and supports the MC (10\,mK) plate.

\subsubsection{The refrigeration system}
%The only way to cool and keep at 10\,mK a huge detector is to use a powerful Dilution Refrigerator. The principle of the dilution of $^{3}$He in $^{4}$He was originally suggested by H. London first in 1951 and later in 1961 together with Clarke and Mendoza~\cite{London1962}, then the first DR was built in the mid ‘60s at the Kamerlingh Onnes Laboratory, in Leiden. Nowadays, the lowest temperature reached by DR is close to 2\,mK, and the most powerful DRs cooling power are of the order of few mW at 100\,mK. 

\begin{figure}[!t]
    \centering
    \includegraphics[width=0.315\textwidth]{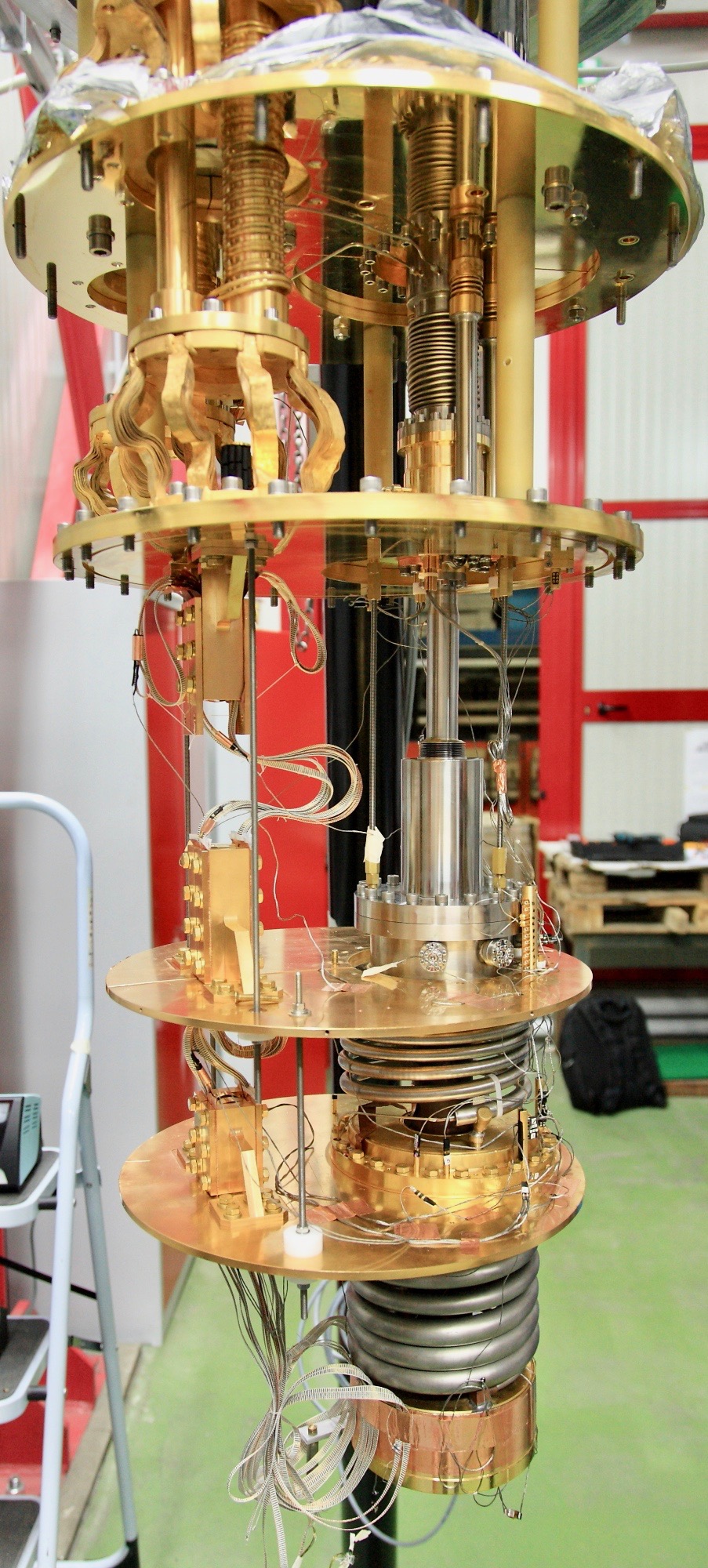}
    \includegraphics[width=0.6\textwidth]{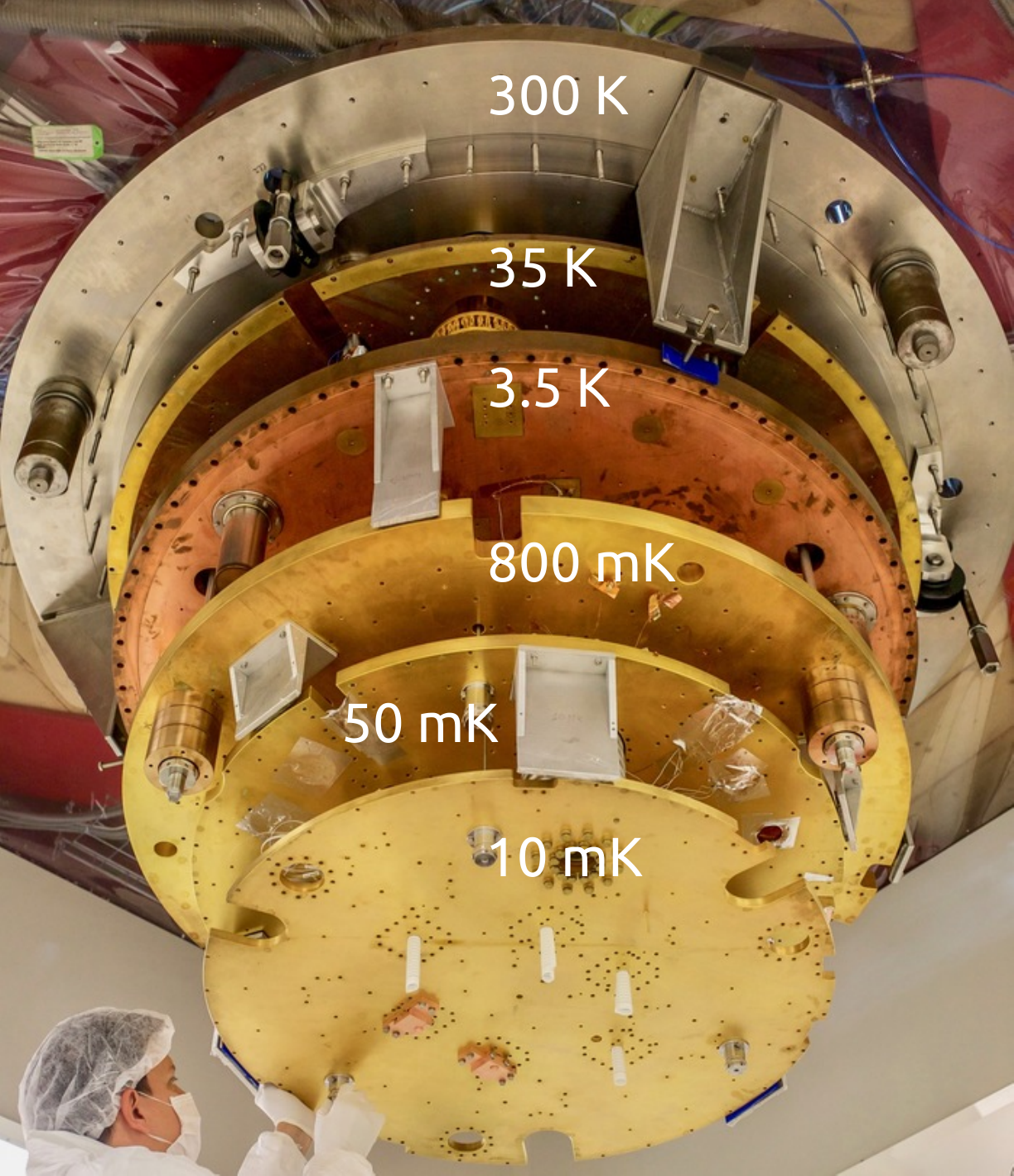}
    \caption{The CUORE Dilution Refrigerator (left side) mounted in the test cryostat and the full CUORE cryostat (right side) with the temperature of the different plates.}
\label{fig:DU1}
\end{figure}

For CUORE, a customized Leiden Cryogenics~\cite{Leidencryogenics} DRS-CF3000 continuous-cycle DR was chosen. It provides cooling power of 4\,$\mu$W @ 10\,mK (about 2\,mW @ 100\,mK), and its base temperature was measured to be 5.5\,mK with no external load. At the time of commissioning, it was the most powerful DR in existence.
It is provided with two condensing lines in parallel for the $^{3}$He/$^{4}$He mixture. The circuit with two independent lines has two advantages: during the cooldown, it allows for increases in the cooling power allowing for flows of $\sim$8\,mmol/s and speeding up the cooldown; and during the steady state circulation, it permits the continued use of the DR in case one of the lines becomes unusable.
The DR was initially tested and commissioned inside a test cryostat, where its working parameters were measured before being mounted inside the CUORE cryostat.
The test cryostat with the DR (Fig.~\ref{fig:DU1}) was also used to test the thermalizations for the CUORE detector wiring, composed of 0.1\,mm NbTi core with 5\,$\mu$m CuNi coating, grouped into 100 woven ribbon cables with 13 twisted pairs in each cable. Dedicated clamps were designed and built in order to thermalize the ribbons from the 4\,K down to the MC plates, and their effectiveness was tested inside the test cryostat.

% The CUORE wiring is a set of 2600 electrical wires used for the crystal bolometers signal and heaters pulse. 

In order to allow $^{3}$He/$^{4}$He mixture condensation, dilution refrigerators can work only in presence of a precooling system at the 4\,K level. For the CUORE cryostat a cryogen free system (see Section~\ref{sec:DilRef}) has been designed and built in order to guarantee high duty cycle runs.
A set of five Cryomech PT415-RM Pulse Tube refrigerators were selected, with nominal power of 1.2\,W @ 4.2\,K and 32\,W @ 45\,K each.

The thermal load of the cryostat was calculated to be 96\,W @ 45\,K, 3.6\,W @ 4.2\,K, 3\,mW at the Still stage, 125\,$\mu$W at the HEX stage, and 4\,$\mu$W at the MC stage. Three PTs were deemed to satisfy the requirements. For redundancy two extra PTs were installed. The reason for the a double redundancy is that in case of a failure of one PT, the thermal input induced from the presence of the spare inactive PT has to be compensated, which is far from negligible (9\,W @ 45\,K and 0.4\,W @ 4\,K). The operating configuration of CUORE is thus to run four PTs all the time keeping one PT switched off as backup. 
In case of a problems with one of the operating PTs, it can be replaced by turning on the backup PT. \ptsnote{Not needed, this is implied.}

As described in section~\ref{sec:Bas_DR}, unlike other mechanical refrigerators, PTs with remote motor option do not have moving parts in the cold head, which is an advantage, both for the reduced vibrational noise and for the diminished magnetic interference. Nevertheless, compressor, rotating valve and helium pressure waves transmit non-negligible mechanical vibrations, so several mechanical decouplings have been implemented. First, a remote motor option was chosen (the rotating valves are separated from the cold head), which also allows for decoupling of the electrical grounding of the cryostat from the external world. Second, a polyurethane ring was mounted on the 300\,K flange and flexible copper braids were inserted between the PT cold fingers and the respective cryostat plates. Moreover, all the external devices close to the cold head (He lines, vessels and rotating valves) were suspended from the ceiling by means of ropes in order to avoid the contact with the cryostat, and He lines pass inside sandboxes before entering the Faraday cage. All these tricks help to minimize vibration transmission.
Operating four PTs at the same time, each one with slightly different rotating frequency (close to 1.4\,Hz), generates mechanical interference patterns, that in some cases can amplify the collective oscillation. To reduce this effect, an active noise cancellation technique~\cite{ActiveNoise} has been developed, driving the PT phases by means of a dedicated software and stepper motors that control the speed of the rotating valves.

\subsubsection{The building and the support structure}
\begin{figure}[!t]
    \centering
    \includegraphics[height=0.5\textwidth]{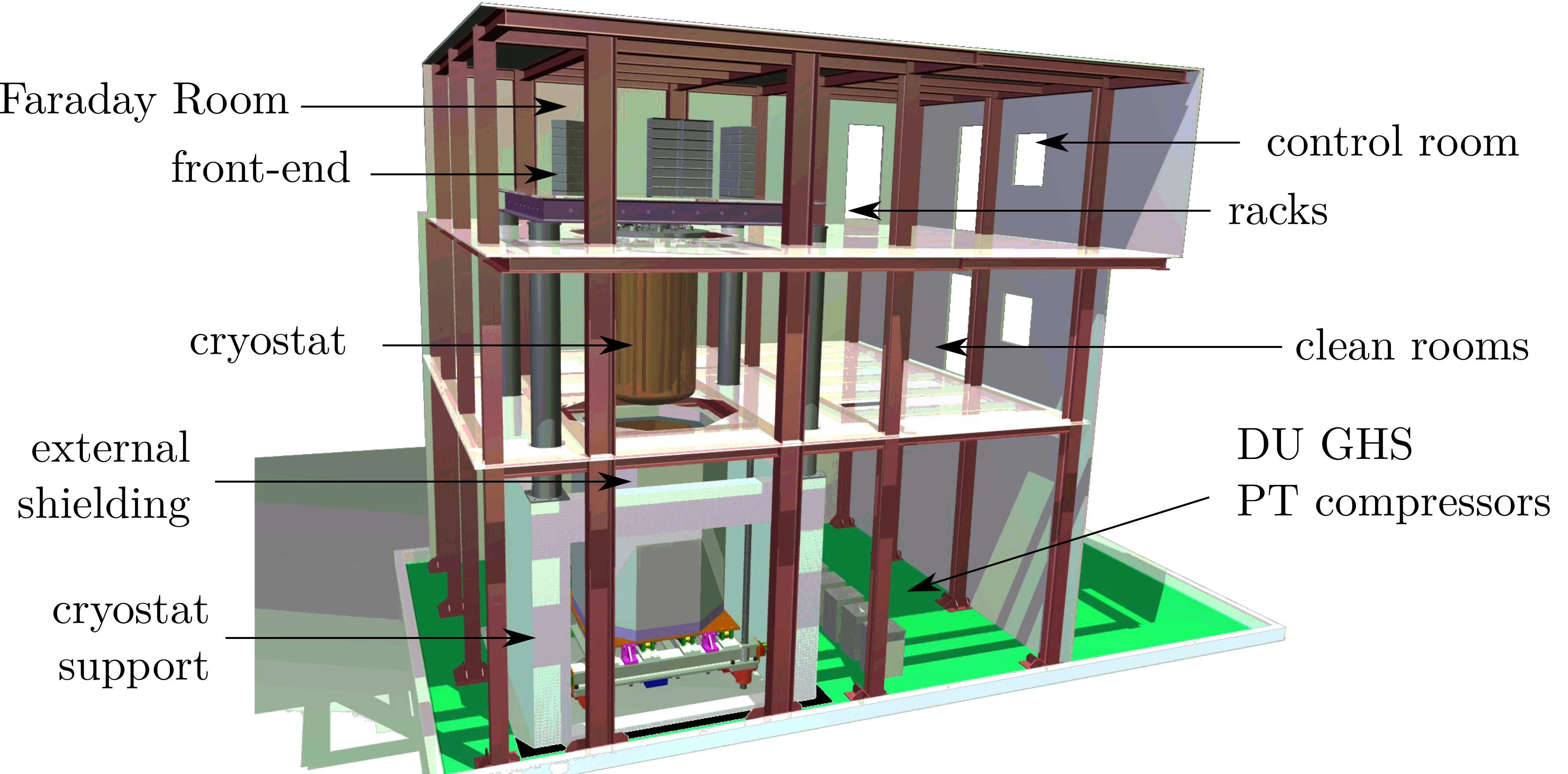}%Hut.png}
    \caption{Rendering of the CUORE building in the Hall A of the LNGS.}
\label{fig:Hut}
\end{figure}

% \paolo{NOTE: in the following the floors are labelled according to  the italian notation: ground, first, and second, but maybe they should be labelled: first, second, and third.}\ptsnote{Commented for now; we can just wait and see if there are complaints about this. I dont' have a strong opinion either way.}
The CUORE experimental setup is hosted in a building (Fig.~\ref{fig:Hut}) located in the Hall A of LNGS. The ground hosts floor the Gas Handling System (GHS) of the DR, the Pulse Tube compressors, the cryostat support, and the external shielding. During the data taking, the external shielding is lifted-up around the cryostat, which is located inside a clean room on the first floor. The cryostat is suspended from the top flange, which is placed on ground of the second floor. The front-end electronic racks are located above the top flange. The space around the top flange is surrounded by the Faraday cage, a 6\,m x 6\,m x 2.5\,m metal wall room, whose purpose is to screen the front-end electronics from electromagnetic fields up to hundreds of MHz~\cite{FARADAY}. Its walls are made of a sandwich of a layer of high permeability metallic glass alloy, called Skudotech, held by two aluminium sheets. The experimental control room is located beside the Faraday cage.

\begin{figure}[!t]
    \centering
    \includegraphics[width=0.6\textwidth]{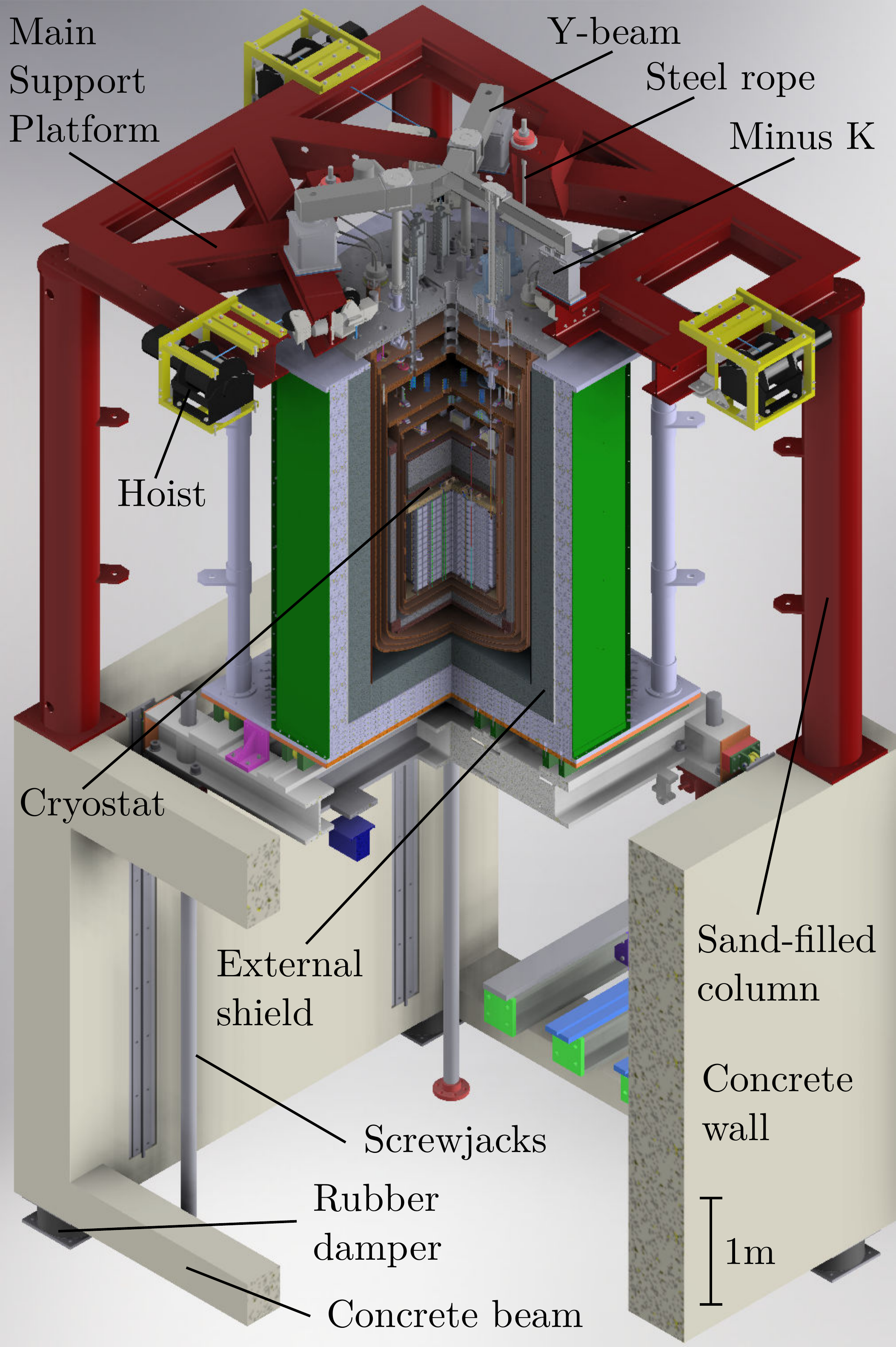}%CryostatStructure2.png}
    \caption{Rendering of the cryostat support structure showing MSP, Y-beam, and Minus-K\textsuperscript{TM} insulation system.}
\label{fig:supportstructure}
\end{figure}

The cryostat is supported by a complex structure~\cite{Alduino:2019xia} that suspends it and, at the same time, isolates the detector from any source of mechanical noise (Fig.~\ref{fig:supportstructure}). To achieve this, the choice was made to support the detector independently from the rest of the cryostat. Great care has been taken to avoid any mechanical vibrations on the crystals, both to reduce the heat caused by microfrictions, and to reduce the microphonic noise experienced by the NTDs --- induced by their high electrical impedance.
 %because at the milliKelvin level even small friction on the detector structure can cause heat
%  , and because the NTDs, due to their high electrical impedance, are extremely sensitive to microphonic noise.
The basement is made of 4.5\,m x 4.5\,m reinforced concrete walls, 0.6\,m thick, connected with the foundation by means of four seismic insulators. Above the wall, four sand-filled steel columns hold up the MSP --- a grid of steel beams that support both the cryostat and the detector support structure. The cryostat is held by 3 ropes connected to the MSP, rigidly anchored to it on the horizontal plane. The detector is connected to a Y-beam, placed above the center of the MSP and anchored to it through a Minus K\textsuperscript{TM} insulation system~\cite{minusk} mechanically decoupling the detector from the rest of the structure.

The external shielding is aimed at protecting the detector from external radioactivity. It is composed of a basement and a lateral shield: on the ground, a pavement made of a 250\,mm thick layer of lead over another 200\,mm thick borated polyethylene layer protects the detector towers from the bottom. Laterally, an octagonal multi-layered shield is installed. Starting from the outside, a 180\,mm thick polyethylene layer surrounds a 20\,mm H$_{3}$BO$_{3}$ panels which in turn enclose a 250\,mm thick lead layer. Lead absorbs environmental $\gamma$ rays, while polyethylene and H$_{3}$BO$_{3}$ are used to thermalize and then capture neutrons~\cite{Alduino2017epjc}.
The total mass of the external shield is approximately 70\,tonnes of lead and 6\,tonnes of polyethylene.

\subsection{Possible future improvements}
\label{sec:possib_futu}
The first three years of operation of CUORE allowed to collect, in addition to a tonne$\cdot$yr of 0$\nu\beta\beta$ data \cite{CUORE_NATURE_2021}, a lot of information on the performance of the cryogenic system and of the operation of detectors in the CUORE facility. These data will allow for improvements to the system in the future in case of a possible upgrade of the CUORE cryogenic system, or in view of the future CUORE upgrades (see for details Section~\ref{sec:future}). From studies performed on cryostat performance, two crucial development paths emerged:
\begin{enumerate}
    \item reducing the sensitivity of the cryostat structure to vibrations and, 
    \item improving the soft thermalization system of the PTs two reduce the vibration transfer.
\end{enumerate} 

One of the main observations during the cryostat commissioning (see Section~\ref{sec:commissioning}) has been that the system is extremely sensitive to external vibrations as well as to vibrations generated by the PTs. To prevent relative oscillations between different parts of the cryostat, a set of stops and constraints have been installed on the bars that connect different plates. A better design and modelling of these vibrational modes will allow us to make the system more rigid avoiding the forced pendulum modes that enhances vibrations.

The second evidence in the study of detectors' performance is the correlation of the noise power spectra with the frequencies excited by the PTs (1.4 Hz and its harmonics). These frequencies are within the range of the CUORE signal bandwidth and hence are particularly detrimental to CUORE data. To minimize the propagation of these frequencies from the PTs to the detectors, the thermal coupling between PTs and the 4\,K and 40\,K plates has been implemented using copper braids. However, a non-negligible amount of vibrations still passes through and several studies are ongoing with the goal to further reduce vibration transfer from PTs. Possible improvements could be made based on developments in the field of gravitational wave interferometers where the problem of vibration suppression is widely studied~\cite{Kagra2005}. 

\section{Commissioning and performance of the CUORE cryostat}
\label{sec:commissioning}

The commissioning of the CUORE cryostat took $\sim$3 years including installation, characterization, and testing of the all the components. It involved the following  steps:
\begin{enumerate}
    \item Commissioning of the DU
    \item Commissioning of the 4K (3.5K) cryostat
    \item Commissioning of the full cryostat
    \item Commissioning of the wiring
    \item Commissioning of the Pb shields
\end{enumerate}
During these commissioning phases a number of cold runs were performed and a variety of problems were uncovered and resolved. For clarity, details of the commissioning are presented based on the above classification instead of chronological order as many of the tests were performed as part of the same cold run. 

In order to monitor the status of the cryostat during the cooldown and characterization, more than 50 thermometers were installed. Given the broad range of temperatures from environmental temperature (300\,K) down to detector base temperature (<10\,mK), several sets of thermometers were used. Temperatures down to a few kelvin were recorded by commercial silicon diode resistance thermometers. For temperatures below 1\,K to a few tens of mK, both commercial and custom-made ruthenium oxide resistor thermometers were used. Finally, on the MC plate, a Magnetic Field Fluctuation Noise Thermometer (MFFT-1) by Magnicon~\cite{Engert_2012}, and a Cerium Magnesium Nitrate (CMN) thermometer~\cite{Greywall_1989}, were used to monitor temperatures down to a few milikelvin. These sensors were calibrated to a Superconducting Fixed Point device also installed on the MC plate. This sensor is based on a set of 9 superconducting transitions spanning from In (3.3 K) to W (15 mK), that allow precise thermometer calibrations. Below the Top Lead, bulk radioactivity places constraints on the sensor choice. Therefore, above a few kelvin, the detector temperature was monitored by ``bare chip'' diode thermometers. Once at base temperature, NTD Ge thermistors were used for both temperature monitoring and for temperature stabilization.

\subsection{Dilution unit commissioning}
\label{subsec:DU}

% The CUORE DU was designed to guarantee the achievement and maintain of cryostat inner stage temperatures, allowing detector operation at 10 mK. To a
Accounting for the different thermal loads at the different temperature stages, the required cooling power for the CUORE DU were estimated to be 3\,mW at the Still stage (800\,mK), 125\,$\mu$W at the HEX stage (50\,mK) and 4\,$\mu$W at the MC stage (10\,mK). The value of 6\,mK in the DU design specifications was deemed an adequate target for the base temperature taking into account the differences in internal arrangement between the test cryostat and the complete CUORE cryostat.
The system was tested at Leiden cryogenics before delivery. At LNGS, the DU was characterized for its cooling power and mixture flow in a dedicated cryostat provided with two CUORE PTs and connected to the gas handling system.
The optimal flow value of $\sim$ (800--1000)~$\mu$mol/s at the base temperature of 5.16\,mK was found by injecting different powers on the Still and monitoring the MC temperature.
By injecting power on the different stages, it was also possible to test the construction specifications of cooling power. The resultant values were found to be 2\,mW at 100\,mK (3\,mW at 123\,mK) and 4\,$\mu$W at 10\,mK.

\subsection{4K (3.5K) cryostat commissioning}
\label{subsec:4K}
\begin{figure}[!t]
    \centering
    \includegraphics[height=0.5\textwidth]{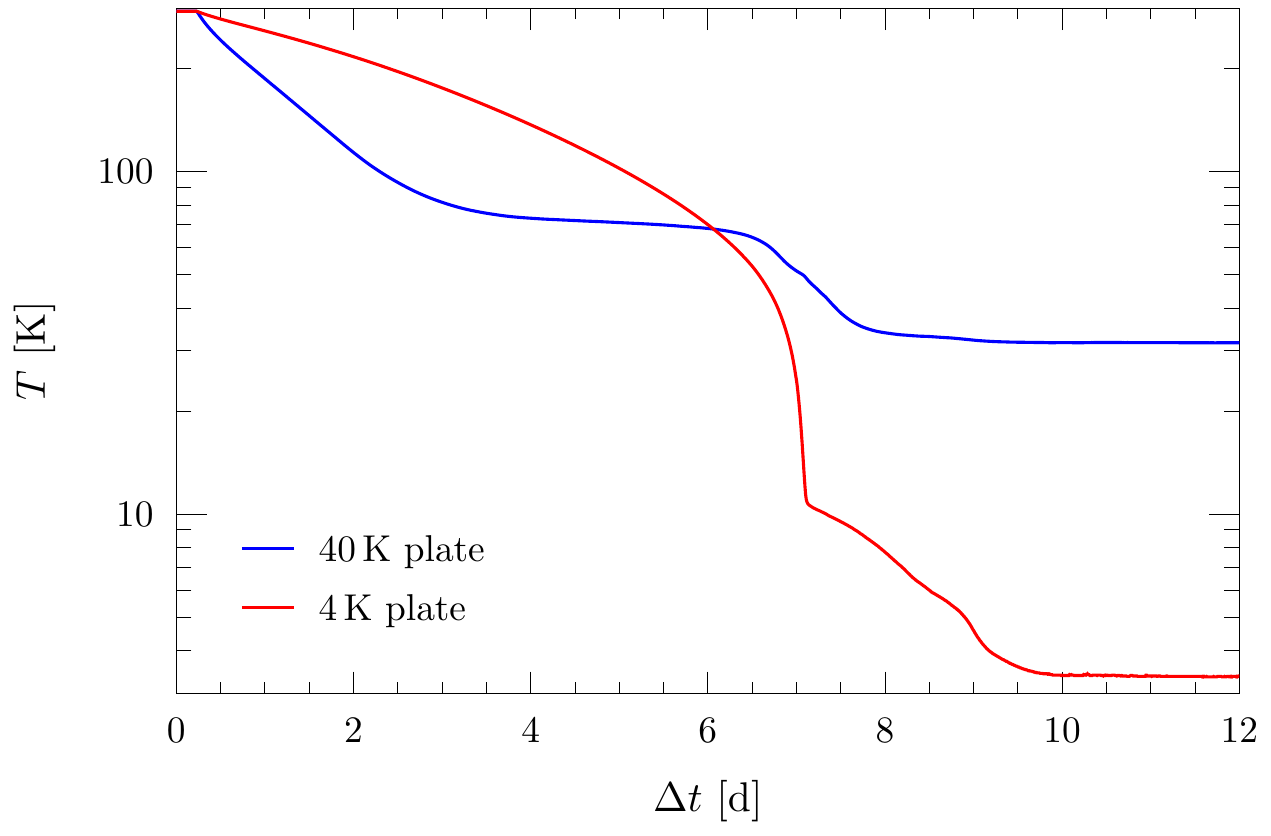}%Hut.png}
    \caption{The temperatures of the 40\,K and 4\,K vessels of the CUORE cryostat during the commissioning of the 4\,K stage \cite{Alduino:2019xia}.}
\label{fig:4Kcom}
\end{figure}

The first step towards the full test of the CUORE cryostat was the assembly and  characterization of the outer cryostat, including the 300\,K, 40\,K, and 4\,K plates and vessels. An extensive leak check campaign was carried out to test all the vacuum seals at the level of OVC (o-rings) and IVC (indium sealings). The cooldown was performed with 3 of the 5 PTs to allow parallel testing of the DU as described in Section~\ref{subsec:DU}. 
As can be seen in Fig.~\ref{fig:4Kcom}, the cooldown took about 12 days. The 40\,K stage cooled down faster than the 4\,K stage due to the higher cooling power of the first stage of PT and because of the the lighter mass to cool at this stage. At about 100\,K, the 40\,K stage started stabilizing and the  4\,K stage temperature crossed the 40\,K stage. The final temperatures of 32\,K and 3.4\,K were reached at the 40\,K and 4\,K stages, respectively. These values are compatible with the nominal cooling power.
Major problems during 4\,K commissioning campaign were related to thermal radiation from OVC plate (300\,K) incident upon the 4\,K stage. Since the cabling for the detectors was not yet installed, all the wire feed-throughs represented open lines of sight from 300K to 4K. To avoid this extra thermal load on the 4K stage, special set of temperature baffles were installed inside the wire feed-throughs during this phase.

\subsection{Full cryostat commissioning}
\label{subsec:full}
This commissioning step consisted of the DU and the 4K system to realize the full CUORE cryostat. This operation included the full installation of the Still, 50 mK (HEX) and MC plates, vessels, the insertion of the DU, and of the two remaining PTs. 

In this long testing campaign, major problems were identified with the following origins: 
\begin{enumerate}
    \item the thermalization of the support bars of the different stages,
    \item the thermal load due to irradiation
\end{enumerate}
As can be seen from Fig.~\ref{fig:bars}, the bars supporting plates are not all in cascading format but for mechanical reasons are connected on higher stages. In particular, the bars supporting the Still plate, that are designed to support the weight of the lateral lead shield, hung directly from the 300\,K steel plate. To minimize power dissipation from 300\,K to 600\,mK the bars were thermalized at different stages. These thermalizations are important and their testing was one of the critical parts of the full cryostat commissioning. 

The problem of thermal irradiation was instead related to the fact that in the``bare'' cryostat (before installation of all the wiring and components), many ports on the different plates represented a line of sight from different temperature stages. A careful work to avoid this problem was carried out, installing caps and baffles to minimize these effects. 

The main difference compared to the 4K commissioning was that, in order to properly cool the lower stages (Still, HEX, and MC) to 4K in the fist part of the cooldown, the gas was injected inside the IVC volume to act as heat exchange gas. In this way the inner stages which are thermally weakly coupled with the 4K stage, were kept at the IVC temperature and reached 4K all together. To avoid helium condensation at 4.2\,K, heat exchange gas must be removed from the IVC above 10\,K.
The final commissioning showed that, with 5 PTs, the base temperature of the 4K stage can be reached in about the same time needed for the IVC and the 40K and 4K stages to reach 34.5 K and 3.9 K, respectively. Once the heat exchange helium gas was completely removed (to a level of 10$^{-9}$\,mbar$\cdot$l/s), the DU operation was started and stable values of 600\,mK, 50\,mK, and 6\,mK were attained in about one day for the Still, HEX and MC stages respectively. The temperature showed a stability of better than 1\% over few days of operation at 6\,mK.

\subsection{Wiring commissioning}
\label{subsec:wires}

Following the major achievement of commissioning the full cryostat, the next step in commissioning of the CUORE cryogenic infrastructure was the installation of the readout wires, running from 300\,K to the MC plate. 
% The detector wiring consists of woven ribbon cables with twisted pairs of NbTi wires in a NOMEXTM texture, with a total of 2600 wires \cite{}. \ptsnote{This was removed since it was already discussed twice before.}
Although the wiring system had already been tested in dedicated setups, it was still crucial to verify the actual impact on the cryogenic performance of the CUORE cryostat.
The wire thermalization down to 4\,K occurs via radiation transmission along the PTFE spirals inside the stainless steel lines. Inside the IVC, ribbons are sandwiched between gold-plated copper clamps installed above and below the Still and HEX plates. On the MC, the junction boxes for the connection between the ribbons and the cabling from the calorimeters act as thermalizers. Measurements of the temperatures, both on the plates and along the wires, showed values in line with the expectations and the cooldown was considered successful.
The installation of readout wires also allowed for the first time to run a calorimetric detector inside the CUORE cryostat: the so called Mini-Tower, a setup equivalent to a 2-floor CUORE tower (8 crystals). The Mini-Tower was mounted directly to the MC plate. The goal of this setup was not to observe calorimetric performance, as the MC plate was expected to be quite noisy in terms of vibrations, but to observe the behavior and performance of NTD sensors for the first time in the CUORE experimental volume.

\subsection{Lead shields commissioning}
\label{subsec:lead}

\begin{figure}[!t]
    \centering
    \includegraphics[width=0.7\textwidth]{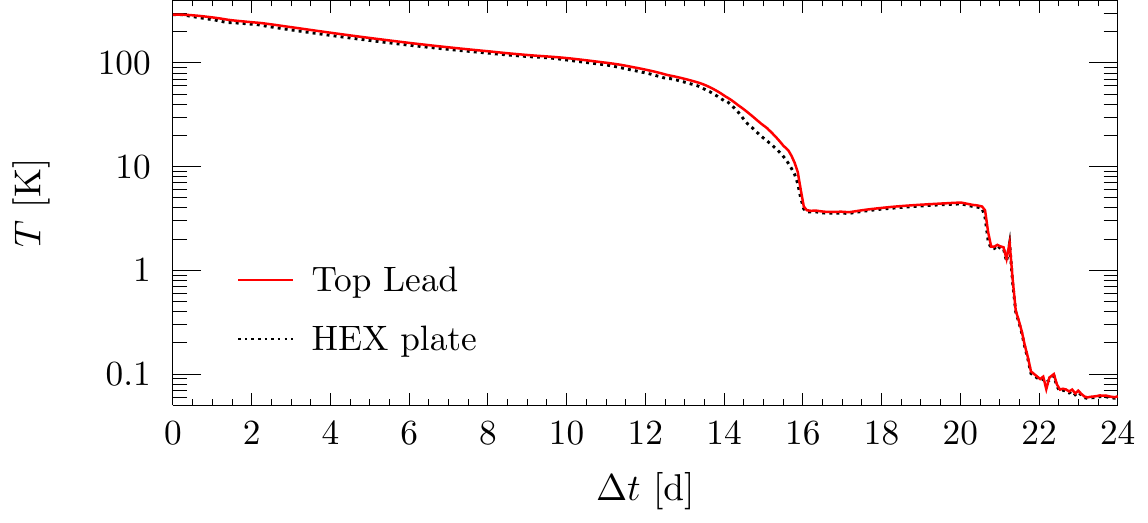}
    \includegraphics[height=0.4\textwidth]{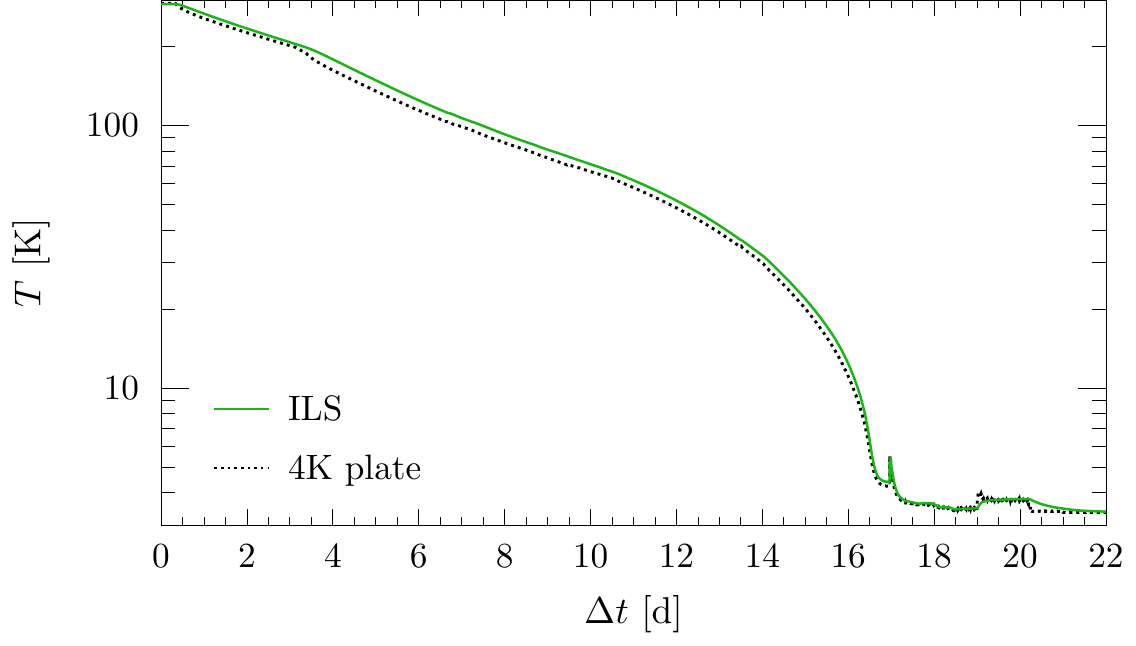}
    \caption{The temperatures Top Lead and  vessels of the CUORE cryostat during the final commissioning runs \cite{Alduino:2019xia}.}
\label{fig:LeadShields}
\end{figure}

The last step of the cryostat commissioning was the installation of the two cold lead shields, the top and lateral lead shields described in section~\ref{sec:cryostat}. These shields, thermalized respectively at 50\,mK and at 4\,K represent a significant mass and thus thermal load for the cooldown.
The shields were installed in two subsequent cold tests allowing for an accurate characterization of their individual contributions to the heat load budget. 

Both the cooldown periods were concluded successfully, as can be seen from Fig. \ref{fig:LeadShields} where the temperature of the Top Lead compared to the HEX stage is reported as well as the temperature of the ILS compared to thes 4K stage. The Mini-Tower was also operated in these conditions, connected to the TSP and thus profiting from the decoupling from cryostat vibrations. For the first time it was possible to observe calorimetric detectors operating inside the CUORE experimental volume. 

These results were considered the final achievements of the CUORE cryostat commissioning and allowed the collaboration to start installing the CUORE detector in the cryostat. Nevertheless cooldown of the detector can be considered the real final run of the commissioning as the cryostat was never operated with a complex object such as the CUORE detector array installed in the experimental volume.

\subsection{CUORE cooldown}
\label{subsec:CUOREcooldown}

The installation of 19 towers and of all the ancillary systems constituting the CUORE detector was performed in a special clean room area with suppressed radon content~\cite{RnMitig}. After the completion of the installation and the channel testing campaign, the cryostat was finally closed and prepared for cooldown.

\begin{figure}[!t]
    \centering
    \includegraphics[height=0.4\textwidth]{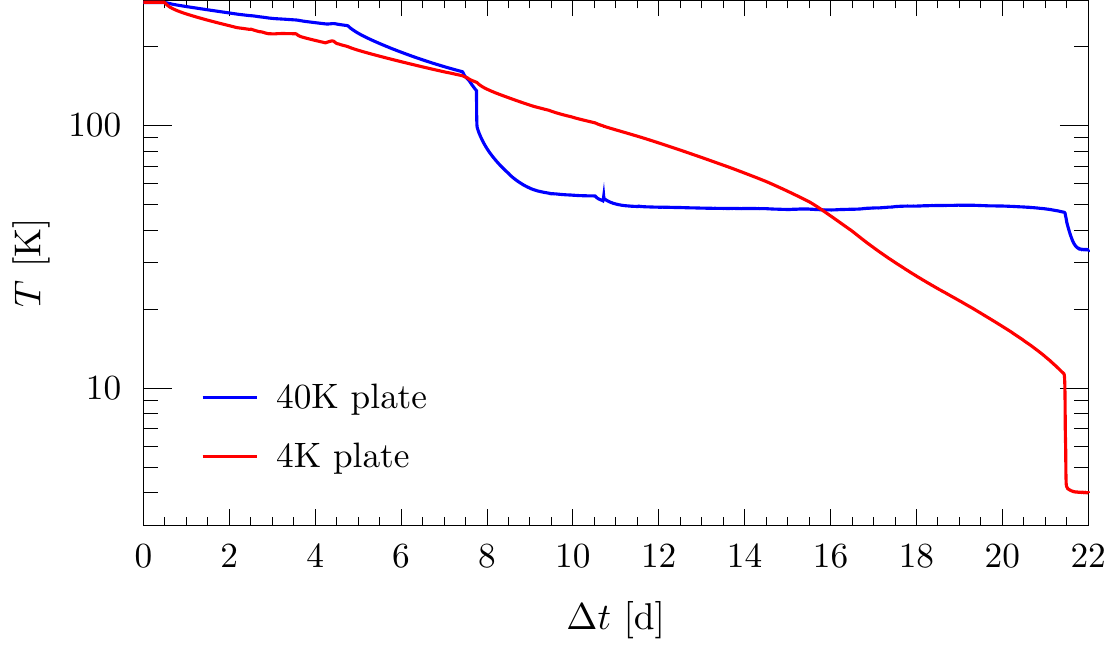}
    \includegraphics[width=0.7\textwidth]{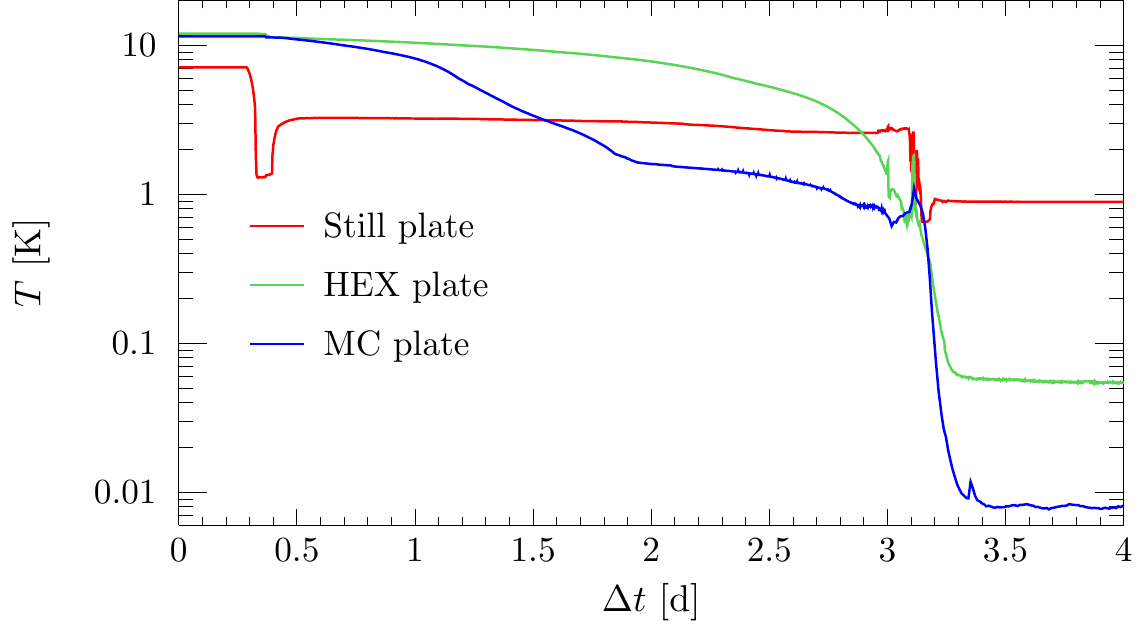}
    \caption{The cooldown of the CUORE cryostat \cite{Alduino:2019xia}. (Top) Temperatures as a function of time of the 40K and 4K stages of the CUORE cryostat after the FCS was turned on. The abrupt changes in the temperature corresponds to the turn-off of FCS at $\sim$150 K. (Bottom) Temperatures of the stages in IVC as a function of time after the DU was tuned on. It can be seen that the base temperature was reached within 4 days of the start of DU circulation.}
\label{fig:CUOREcoold}
\end{figure}

The progression of the CUORE cooldown is shown in Fig. \ref{fig:CUOREcoold}. The first part of the cooldown  was supported by an auxiliary helium gas precooling system, named Fast Cooling System (FCS) \cite{pagliarone2017cuore} to lower the stress on the PTs, injecting cold helium gas in the IVC volume. The FCS was then turned off and disconnected once the thermal stages reached 150\,K. As expected, the cooldown took 20 days before turning on the DU unit. The cooldown was paused for about one month to perform intensive pumping of the IVC volume to remove the residual heat exchange gas (to a level of 10$^{-7}$\,mbar$\cdot$l/s).  Once the DU was turned on, it took less than 4 days to reach the base temperature of 8\,mK on the MC, with 0.89\,K and 55\,mK on the Still and HEX stages respectively. The temperature of 5.9\,mK was initially reached in the experimental volume for the full detector.

This result represents the starting point of the CUORE science runs which in the first 3 years, after the setup and optimization campaign, already provided many relevant scientific results \cite{CUORE_PRL_2018,CUORE_PRL_2020,CUORE_NATURE_2021,Adams:2021xiz,Alduino:2017mnz}. Moreover the success of the CUORE cryogenic infrastructure opens the way to new generations of cryogenic experiments, including a possible CUORE upgrade in the same cryogenic facility (CUPID) and a new generation of mK large infrastructures capable of hosting tonne or multi-tonne cryogenic experiments.

\section{Implications for future applications}\label{sec:future}
In this section an overview of the possible future applications of the cryogenic technology pioneered by the CUORE cryostat is given. Many different fields with diverse applications will benefit from the ground-breaking work by the CUORE experiment. 

In the first part, the natural upgrade of the CUORE infrastructure to host the CUPID project is presented, highlighting the prospect of improved sensitivity to 0$\nu\beta\beta$.
Applications of tonne-scale cryogenic facilities in other fields are described in the subsequent sections. DM searches, in particular focusing on low mass DM are discussed. %, presenting also the possible CUPID sensitivity to this search.
Search for axions, observation of supernova neutrinos and measurement of coherent elastic neutrino-nucleus scattering all require dedicated facilities with tonnes of active detector mass to be competitive with the next generation of experiments.
Finally the advantage of using large cryogenic infrastructures in one of the most active fields of applied physics, the development of quantum computing, is presented.

\subsection{Neutrinoless Double Beta Decay: the CUORE Upgrade with Particle IDentification}
\label{sec:Impl_ndbd}

The results of the CUORE experiment convincingly proved the potential of cryogenic calorimeters in terms of energy resolution, efficiency and reproducibility. 
This success sets the basis for a next generation experiment, CUPID. CUPID will exploit CUORE's unique cryogenic facility to deploy a new detector, with the primary goal of reaching sufficient sensitivity to fully cover the region of Majorana mass corresponding to the inverted neutrino masses hierarchy.
To this end, CUPID will increase the number of $\beta\beta$ decays through isotopic enrichment and suppress the total background by two orders of magnitude, obtaining a background-free window around the 0$\nu\beta\beta$ decay Q-value~\cite{CUPID_preCDR_2019}.

The implementation of a background free experiment hinges on two major improvements with respect to CUORE.
The first is that CUPID will rely on calorimeters that, unlike TeO$_2$, emit scintillation light at cryogenic temperatures. 
The simultaneous readout of the calorimetric and scintillation light signals enables particle identification and thus the rejection of dominant backgrounds, primarily composed of $\alpha$ particles. 
Second, the choice of an isotope with a Q-value higher than the 2.6\,MeV line of $^{208}$Tl, will place the CUPID ROI above most of the natural $\beta/\gamma$ radioactivity.
The potential of this approach was confirmed by two medium scale experiments, CUPID-0~\cite{Azzolini_2018,Azzolini_2019} and CUPID-Mo~\cite{Armengaud_2020,Armengaud_2021}, which demonstrated a background in the region of interest for neutrinoless double beta decay search at the level of 10$^{-3}$\,counts/keV/kg/yr.

The 0$\nu\beta\beta$ isotope selected by the CUPID collaboration is $^{100}$Mo, with a Q-value = 3034.40 $\pm$ 0.17\,keV~\cite{RAHAMAN2008111}. 
Following the successful validation with the CUPID-Mo demonstrator, natural molybdenum will be highly enriched in $^{100}$Mo to overcome the relatively low natural isotopic abundance of this isotope of 9.74\%~\cite{meija2016isotopic}.
The enriched material will be used to synthesise Li$_2^{100}$MoO$_4$ scintillating crystals~\cite{Cardani_2013,Armengaud_2017}, to be operated as cryogenic calorimeters in an array featuring a total mass of about 472\,kg (corresponding to $\sim$253\,kg of $^{100}$Mo).

Each Li$_2^{100}$MoO$_4$ crystal will be coupled to a light detector. 
Since standard technologies for light detection, such as photomultipliers or photodiodes, do not work properly at 10\,mK, CUPID will use other cryogenic calorimeters as light detectors~\cite{Beeman_2013}: germanium wafers, which demonstrated excellent performance in the CUPID-0 and CUPID-Mo demonstrators.  
The Ge wafers will be cut to a thickness of hundreds of $\mu$m, significantly reducing the heat capacity of the device thus allowing the measurement of much smaller heat signals. 
This is needed since the signal emitted in the form of scintillation light is only a small fraction of the total interaction energy, only a few keV of light emitted by a 3 MeV electron.
As both the Li$_2^{100}$MoO$_4$ and the Ge wafers will be equipped with NTD Ge thermistors, it will be possible to utilize the same readout technology developed for CUORE with  minor modifications.

With the Li$_2^{100}$MoO$_4$ technology well established, demonstrating an energy resolution of $\sim$5\,keV FWHM~\cite{Armengaud_2020} in the region of interest, ultra-low contamination levels ($<$3\,$\mu$Bq/kg both in thorium and uranium) and high reproducibility, the scintillating calorimeteric technique is ready to be expanded to a tonne-scale mass.
Several activities are ongoing to finalize the detector design~\cite{armatol2020characterization, Armatol_2021}.
The CUORE cryostat is the natural choice to host the CUPID detector. 
The required upgrades to the cryogenic facility include the suppression of mechanical noise and increasing the number of available readout channels, while maintaining the same cooling power. 
While increasing the number of readout channels can be a relatively straight forward process, the suppression of mechanical noise demands some modifications of the cryogenic facility, as described in section~\ref{sec:possib_futu}.
These upgrades are being tested on separate cryostats and will be ported to the CUORE cryostat when the CUORE experiment concludes physics program. 

Aside from the dilution refrigerator, the CUORE infrastructure also offers many complementary facilities that would significantly contribute to the deployment of a next-generation experiment. 
The detector assembly lines, the clean rooms built around the dilution fridge, the radon-free system, among other tools developed for the CUORE infrastructure, will play an essential role in building and commissioning future detectors for 0$\nu\beta\beta$ searches.

\subsection{Dark Matter Searches}
\label{sec:dm}
The evidence for dark matter is overwhelming in many cosmological observations at small and large scales, from galactic rotation curves to the interpretation of anisotropies in the cosmic microwave background~(CMB). 
The evidence so far relies on gravitational effects and
despite the huge technological progress in the field of DM search in recent decades, the particle nature of DM still remains unknown~\cite{Tanabashi:2018oca}.
Among the vast number of DM candidates proposed, the WIMP paradigm has dominated over the past few decades~\cite{Roszkowski:2017nbc} thanks to the remarkable coincidence of two unrelated facts: WIMPs not only arise spontaneously in Beyond the Standard Model (BSM) theories,  but also their relic abundance determined from the freeze-out mechanism matches the measured DM density for reasonable ranges of weak-scale annihilation cross-section and WIMP masses ($m_W$) in the range of 1~GeV - 100~TeV. 
Nowadays the approach to the field is more open and the effort to scrutinize much wider regions of the DM parameter space is carried out to constraint a variety of different DM candidates. The main experimental channel still remains the observation of DM particles from the dark halo of our galaxy may scattering off nuclei in a terrestrial detector~(direct detection). 

At present, large regions of the WIMP parameter space for $m_W$ above 10~GeV are excluded by very sensitive experiments based on Xe and Ar dual-phase (gas and liquid) noble gas TPC experiments~\cite{Aprile:2018dbl, Akerib:2016vxi, Wang:2020coa,Agnes:2018fwg}. 
These technologies have achieved incredibly low backgrounds of the order of few counts/tonne/year thanks to the high radiopurity of the target, event-by-event particle discrimination,and 3-D position reconstruction of the interaction site. 
However, the ultimate background in the search for DM comes from coherent elastic neutrino-nucleus scattering~(see section~\ref{sec:cenns}), as solar pp and $^7$Be neutrinos ($m_W<$10~GeV) and atmospheric and diffuse supernova neutrino backgrounds (DSNB) ($m_W>$10~GeV) produce nuclear recoil signals indistinguishable from DM interactions.
Projected extensions of noble TPC experiments are designed to reach so called neutrino floor --- which is the threshold below which neutrino events dominate the DM events --- in the next decade and new technologies are required to further improve the sensitivity. 

%WIMPs in the region of a few GeV also called light WIMPs on the other hand did not get so much attention in past because this mass range was excluded in most stringent models. 
%In addition, to explore this region detectors consisting of light nuclei and very low energy thresholds are required, in order to get an observable energy transfer from the light DM to the nucleus. 
%Work is in progress at present to lower the detector thresholds in order to explore this region. \ptsnote{Need citation}
%. Cryogenic detectors have demonstrated in the past decade that thresholds as low as few hundreds of eV are at reach, so they are leading the search in the mass range.  

The WIMP paradigm remains of great interest, but at the same time --- due in part to the lack of signals in direct, indirect --- or collider searches,  there is a strong motivation to explore a broader set of dark matter candidates.
In particular, if new low-mass mediators are hypothesized, then a keV-to-GeV (light DM) window is open. 
The traditional direct detection approach (search for a nucleus recoil) is impractical as the energy transfer to the nucleus from such light particles is below any reasonable detector threshold. 
However, sub-GeV DM scattering off electrons are still observable.
Novel detector technologies and materials to lower the experimental thresholds further, reaching the single electron level, are currently under development. Many of these technologies are based on cryogenic detection techniques~\cite{Battaglieri:2017aum}.

Cryogenic calorimeters are excellent detectors for light WIMPs or more exotic light DM models, presenting remarkable characteristics: 
\begin{enumerate}
    \item very low energy thresholds: TeO$_2$ large crystals ($\sim$750~g) used in the CUORE demonstrators featured energy thresholds around [10-30]~keV~\cite{Alduino:2017xpk} down to  3~keV~\cite{Alessandria:2012ha}. 
    Nevertheless, the feasibility of reaching thresholds of the order of tens of eV in cryogenic experiments operating smaller crystals has been largely demonstrated by CRESST~\cite{Abdelhameed:2019hmk}, EDELWEISS~\cite{Arnaud:2020svb}, and  SuperCDMS~\cite{Agnese:2018col}
    \item high quenching factor: the electron/nuclear recoil relative efficiency factor (the so called quenching factor) is generally close to unity for heat signals. 
    This parameter --- which quantifies the conversion efficiency  of a nuclear recoil energy deposition into a measurable signal with respect to an electron recoil --- is typically lower than 50\% for ionization or scintillation detectors. 
    The precise knowledge of the quenching factor and its dependence on energy is important since it is used in detector characterization and is a source of systematic uncertainty in  DM searches, since the experimental spectrum is calibrated using $\gamma$ sources and therefore given in electron-recoil equivalent (eVee) units
    \item wide choice of materials: there exists a wide choice of materials that can be used as cryogenic calorimeters, allowing for the exploration of a wide region of the DM parameter space
    \item ease of dual readout implementation: dual readout is easy to implement for semiconductors (ionization) or scintillating crystals (light). 
    The secondary signal is differently quenched with respect to the heat signal depending on the nature of the interacting particle, so event-by-event background rejection is possible
\end{enumerate} 

A large cryogenic facility like the CUORE cryostat offers interesting possibilities in the near future for probing yet unexplored DM parameter regions as well as light WIMPs and sub-GeV DM, both with traditional cryogenic calorimeters as well as emerging technologies that need mK environment.

%%%%%%%%%%%%%%%%%%%%%%%%%%%%%%%%%%%%%%%%%%%%%%%%%%%
%%%%%%%%%%%%%%%%%%%%%%%%%%%%%%%%%%%%%%%%%%%%%%%%%%%
\subsubsection{Search for Light WIMPs with CUPID}
\label{subsec:cupidWimp}
Traditional direct detection experiments search for nuclear recoils caused by DM elastic scattering.
%The average energy transferred in such interaction is gven by $E_{nr}=q^2/2m_N$, where $m_N$ is the mass of the nucleus, $q\propto m_W v$ is the transferred momentum and $v\sim$220~km/s is the WIMP average velocity with respect to the earth. 
%In more intuitive units,$E_{nr}\sim$1~eV$\times(m_W$/100~MeV)$^2\times($10~GeV/$m_N)$. \ptsnote{Is this needed ?}
For scattering in the non-relativistic limit the only two types of possible interactions are (i) scalar, or  spin-independent~(SI), which are coherent interactions of DM with the entire nucleus and (ii) axial, or spin-dependent~(SD), in which the DM interacts with the spin-content of the nucleus. 
SI greatly enhanced relative to SD, with an expected rate proportional to the square of target nucleus mass. 
Furthermore, for spin-zero isotopes only the SI interaction is possible.
For these reasons, SI interaction has been the main focus of experimental efforts, favoring the use of heavy target-nuclei.
%For SD there is no $A^2$ dependence, 
%so in this case light nucleus are preferred.
However, determining of the nature of DM will require the quantification of both types of interactions, since there is no preference for the scattering to occur via scalar or axial interactions. 
%The local DM density is $\sim\rho_W$=0.4~GeV/cm$^2$ 
%with a RMS velocity of about
%270~km/s. This implies a flux at the Earth of 10$^8$-10$^{10}$~s$^{-1}$cm$^{-2}$ for a WIMP of mass $\sim$100~GeV. The
%DM interaction rate in the detector is proportional to the number of target nuclei, the number density of WIMPs
%$\rho_W$/m$_W$, their relative velocity with respect to the detector and the interaction cross-section, which in general depends on the WIMP energy and the transferred momentum, and  for most standard WIMP scenarios are of the order of 10$^{-45}$ - 10$^{-50}$~cm$^2$ and below. 

% The CUPID experiment will operate 472~kg of Li$_2$MoO$_4$ scintillating crystals  as cryogenic calorimeters read by NTD Ge thermistors. 
% Li$_2$MoO$_4$ scintillates at mK temperatures, providing a secondary readout that will be detected by Ge wafers, also equipped with NTD-Ge thermistors. 
The target material in CUPID is a combination of heavy~(Mo) and light elements~(O, Li), the latter resulting in a good sensitivity for light WIMPs. 
Although the search for DM is not the primary scope of CUPID, competitive limits are foreseen in  the light-WIMP region for SI interactions, provided a threshold of the order of 5~keV or better is attained. 
Moreover, CUPID will be an excellent probe for SD-interacting WIMPs, thanks to the large amount of $^7$Li (92.41\% natural abundance) in the target.
This isotope is the best choice among the isotopes with nuclear angular momentum J$_N\neq$0: it is very light, which tends to increase the energy transfer, and has a near-to-maximal value for the total spin of protons averaged over all nucleons $\langle$S$_p\rangle\sim$0.5. 
Thanks to its large mass, depending on the final background and threshold achieved CUPID could lead the proton-only SD search region for light WIMPs in the near future.
% It is also worth mentioning that the thin Ge wafers that work as light detectors in the CUPID setup can also be DM detectors on their own. 
 % Despite their low mass of the order of several grams, energy thresholds as low as few tens of eV are achievable, making them excellent probes for light DM. 
 
\subsubsection{Axion Searches}
The axion was originally proposed as a mechanism to solve the strong CP problem: CP is not expected to be conserved in Quantum Chromodynamics (QCD). 
The violating term depends on an angle $\theta_{QCD}$ which can have arbitrary values, however observations indicate that it is very close to 0. 
The best constraint comes from the experimental upper limit on the neutron electric dipole moment~\cite{Afach:2015sja}: \[|d_n|\sim 3\times 10^{-16 }\theta_{QCD} e\cdot\,cm<3\times10^{-26}e\cdot\,cm~(90\% CL)\]
Robert Peccei and Helen Quinn realised that the strong CP problem could be solved by introducing a global U(1) symmetry (PQ symmetry) which is spontaneously broken~\cite{Peccei:1977hh} and whose associated Nambu-Goldstone boson is called the axion~\cite{Wilczek:1991jgb}. %It acts as ``dynamic'' $\theta_{QCD}$: once the symmetry is broken, the CP violation term dynamically relaxes to zero. 
The properties of the axion depend mainly on the value of the axion decay constant $f_a$, related to the energy scale of the spontaneous breaking of the PQ symmetry. 
The couplings to SM particles are very weak, the most relevant of those being to two photons
($g_{a\gamma}$, which is responsible for the Primakoff effect), nucleons ($g_{aN}$) and electrons ($g_{ae}$), although the last coupling is strongly suppressed in hadronic models.
The main detection mechanism arises from the Primakoff effect in which an axion is converted into a photon in the presence of a strong magnetic field.
Depending on the model, axions can also interact with electrons in atomic shells, in the so called axio-electric and Compton-like effects, or undergo resonant absorption or emission in magnetic nuclear transitions. 
For QCD axions, mass and interaction strength are proportional to each other, forming a diagonal band in the parameter space (mass, coupling). 

Unrelated to the strong CP problem, BSM theories predict the existence of light pseudo-scalar particles with very weak couplings to the SM particles, coming from other spontaneously broken symmetries at high energy scales. 
These particles, generally referred to as axion-like particles (ALPs), do not follow any \textit{a priori} relationship between mass and coupling strength, so candidates cover a wider mass range.

\ptsnote{This whole paragraph may be removed}
The axion is also a DM candidate. Axions and ALPs would have been produced thermally by interactions with SM particles in the early Universe, but their main production mechanism is non-thermal. The typical mass range can span between $10^{-6}$ eV and $10^{-1}$~eV, depending on the generation mechanism.
%When the symmetry breaking occurs before inflation, during the relaxation of the axion field it makes small amplitude oscillations around the minimum of its potential (vacuum realignement), producing axions with very small velocity dispersion. Therefore, in the pre-inflation scenario, axions and ALPs are good CDM candidates provided their mass is in the  $10^{-6} - 10^{-4}$~eV range, in order to produce the right DM density. Depending on the model, masses up to $10^{-3}$~eV can be accommodated, or one can go to much lower masses claiming anthropic arguments~\cite{Irastorza:2018dyq}.
%If the transition happens after inflation, 
%topological defects appear, like axion strings and domain walls, which eventually
%decay producing cold axions, but with 
%larger velocity dispersion than in the pre-inflation scenario. Some studies point to mass values 
%of the order of $10^{-5}-10^{-4}$~eV~\cite{Borsanyi:2016ksw, Dine:2017swf}, 
%but in general calculations are affected by large uncertainties and masses up to $10^{-4}- 10^{-1}$~eV are predicted by some models~\cite{Kawasaki:2014sqa}.

%Main constrains of the axion parameters come form Stellar evolution , as axions provide an additional cooling mechanism, and 
%other astrophysical arguments.
%A recent combined analysis of the M5 tip and WD data gives 
% gae < 2.6e-13  95\%C.L. ref 79 redondo

Axions can be produced in the laboratory by using a photon source and intense magnetic fields, but given the weakness of the interaction, the most promising detection strategies rely on two natural high-intensity sources: the the galactic dark matter halo and the Sun~\cite{Sikivie:1983ip}.
There has been a significant surge in experiments searching for dark matter axions in the galactic halo in the recent past~(see \cite{Irastorza:2018dyq} and references therein). 
Classically, they are performed through axion haloscopes~\cite{Sikivie:1983ip} that search for a resonant signal in a high-quality factor microwave cavity at the frequency corresponding to the axion mass.
In a scenario where the DM halo contains massive ($~\mathcal{O}$(keV)) ALPs with non-zero $g_{ae}$ coupling, they can produce an observable signal (a peak at the energy corresponding to the ALP mass) in direct detection DM experiments or other large-mass low-background detectors looking for rare events.  
Some experiments like CDMS~\cite{Ahmed:2009ht}, EDELWEISS~\cite{Armengaud:2013rta}, or XENON100~\cite{Aprile:2014eoa} have reported limits excluding DM ALPs with $g_{ae}<10^{-12}$, but masses $<$ 1~keV are not reachable. 
A tonne-scale experiment like CUORE or CUPID could improve this limit considerably, but this scenario is not theoretically well motivated. 
New proposals are emerging however based on new detection ideas, such as superconducting LC circuits~\cite{Sikivie:2013laa} or atomic transitions~\cite{Sikivie:2014lha}, which could profit from a huge cryogenic infrastructure like the CUORE cryostat.

Solar axions/ALPs are a more appealing scenario for a tonne-scale cryogenic detector.
They are produced in large quantities in the Sun's core through the Primakoff mechanism, with an energy spectrum ranging from 1 to 10~keV and peaking at 3~keV~\cite{Irastorza:2018dyq}. 
If axions couple to electrons (non-hadronic models), the flux at about 1~keV is dominated by ABC processes (Axio-recombination, Bremsstrahlung and Compton).
In addition, nearly monochromatic axions are emitted in magnetic nuclear transitions via axion-nucleon coupling.

The hunt for Solar axions is being carried out by means of axion helioscopes: x-ray detectors coupled to a powerful magnet in which the axion-photon conversion takes place. 
The most sensitive instrument of this kind is CAST, which reached an experimental upper limit of $g_{a\gamma} < 0.66\times 10^{-10}$~GeV$^{-1}$ for axions masses lower than 0.02 eV~\cite{Anastassopoulos:2017ftl}, and its projected upgrade IAXO~\cite{Armengaud:2014gea}.

Solar axions can also be searched for with crystalline detectors, as axion-photon conversion can also take place in the crystalline electric field. 
The periodicity of the crystal lattice coherently enhances the conversion probability when the Bragg condition is satisfied (i.e., when the  direction of the incoming axion matches one Bragg angle). 
As the crystal lattice orientation with respect to the solar flux changes continuously due to the earth's rotation, this effect produces characteristic sharp peaks in the counting rate both in the energy and time spectrum, clearly distinguishable from other backgrounds.
This method has been applied in some rare-event experiments like COSME~\cite{Morales:2001we}, DAMA~\cite{Bernabei:2001ny}, EDELWEISS-II~\cite{Armengaud:2013rta} and CDMS~\cite{Ahmed:2009ht}, DAMA among them producing the tightest bound so far of $g_{a\gamma}<1.7\times10^-9$~GeV$^{-1}$.
A cryogenic calorimetric experiment with large mass, very low energy threshold and good energy resolution would improve this limit, as discussed in~\cite{Li:2015tsa} for CUORE.
In this work it was shown that an upper limit of $g_{a\gamma}<3.83\times10^{-10}$~GeV$^{-1}$ at 95\%~C.L. can be obtained, provided a good energy resolution in the energy range from 2--6~keV is achieved. 
This bound is independent on the axion mass, and therefore would improve upon the CAST limit for masses above 1~eV.
 
Assuming that axions couple to electrons, a new detection window opens for large mass experiments devoted to rare searches via the axio-electric effect, which is equivalent to the photoelectric effect but with the absorption of an axion instead of a photon. 
%Depending on the solar axion production mechanism the experiment is sensitive to, t
This technique can only produce combined limits in the product of two coupling constants, i.e. $g_{ae}g_{a\gamma}$ for Primakoff axions or $g_{ae}g_{aN}$ for axions produced in a nuclear transition.
A number of searches of this kind have been performed in recent years, see for example EDELWEISS-II~\cite{Armengaud:2013rta}, XMASS~\cite{Abe:2012ut}, XENON100~\cite{Aprile:2014eoa}, PANDA-X~\cite{Fu:2017lfc}, LUX~\cite{Akerib:2017uem}, CDEX~\cite{Wang:2019wwo}, and COSINE-100~\cite{Adhikari:2019tgv}.
Very recently the XENON1T collaboration has reported an excess of events in the electronic recoil spectrum below 7~keV~\cite{Aprile:2020tmw}. 
The excess is compatible with solar axions with coupling constants inscribed in the cuboid defined by $g_{ae}<3.7\times10^{-12}$, $g_{ae}g_{an}^{eff}<4.6\times10^{-18}$ and $g_{ae}g_{a\gamma}<7.6\times10^{-22}$~GeV$^{-1}$. 
The result is very exciting as it is the first hint of a positive axion detection. 
However, as the authors admit, the excess could be from the $\beta$ decays of a tritium contamination in the Xe detector.
A tonne-scale cryogenic calorimetric detector with low enough energy threshold ($\leq$5~keV) and low backgrounds ($\leq$0.1~c/keV/kg/y) could help in solving this controversy, as it could disentangle the dominant ABC axion peaks (monochromatic lines coming from axio-recombination) from Primakoff spectrum thanks to its superior energy resolution.

Finally, the search for monoenergetic axions which originate in a M1 nuclear transition in the sun has been applied for a number of nuclides, e.g. $^{57}$Fe~\cite{Moriyama:1995bz, Alessandria:2012mt, Armengaud:2013rta, Krcmar:1998xn,Andriamonje:2009dx}, $^7$Li~\cite{Krcmar:2001si}, $^{169}$Tm~\cite{Derbin:2009jw}, or the $p+d\rightarrow^3$He+a(5.5~MeV) reaction~\cite{Bellini:2012kz,Derbin:2013zba}.
This technique is especially sensitive when the experimental target contains the same nuclides as the source of the reaction, giving rise to a resonant absorption. 
Therefore the $^7$Li channel will be especially promising in the  Li$_2$MoO$_4$ scintillating crystals of CUPID.

\subsection{Coherent Elastic Neutrino-Nucleus Scattering}
\label{sec:cenns}
Coherent elastic neutrino-nucleus scattering~(CE$\nu$NS) is a neutral current process mediated by the Z-boson.

\begin{figure}[!t]
\centering
\includegraphics[width=0.7\textwidth]{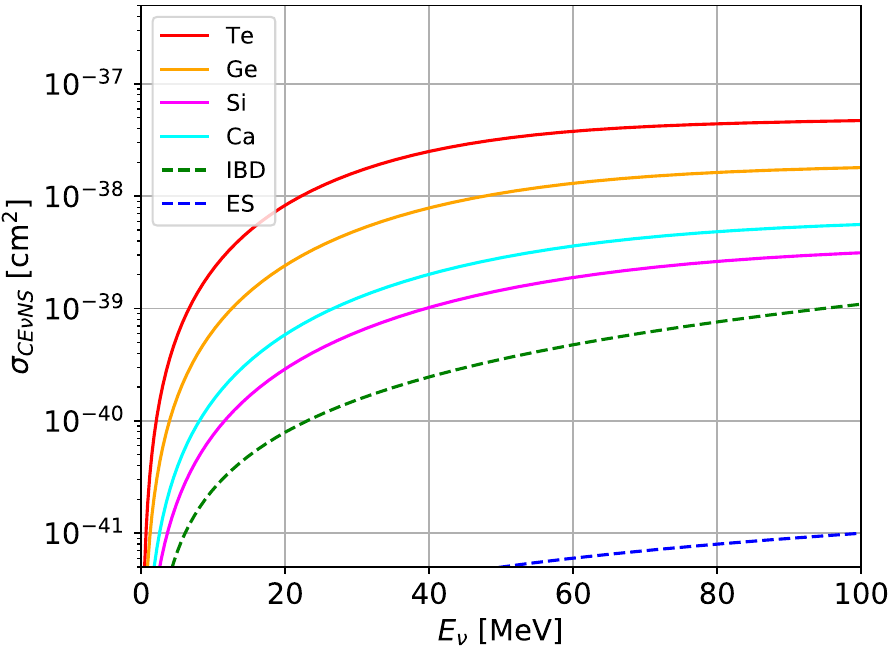}
\caption{CE$\nu$NS cross-section as a function of the neutrino energy for different target nuclei. For comparison electron scattering and IBD interaction cross-sections. Adapted from ~\cite{Pattavina:2020cqc}}
\label{fig:xsec}
\end{figure}

The total CE$\nu$NS cross-section can be easily computed from Standard Model basic principles~\cite{Freedman:1977xn}:
\begin{equation}
\label{eq:xsec}
\frac{d\sigma}{d E_R} = \frac{G^2_F m_N}{4 \pi} Q^2_W(N) \left(1- \frac{E_R m_N}{2 E^2_{\nu}}\right) \cdot F(q)^2
\end{equation}

where G$_F$ is the Fermi’s coupling constant, Q$_W(N)$ the Weinberg angle, $N$ the neutron number of the target nucleus, m$_N$ its mass, E$_{\nu}$ the total energy of the incoming neutrino, and E$_R$ the recoil energy of the nucleus. 
The last term of the equation is the elastic nuclear form factor, $F(q)$, at momentum transfer $q =\sqrt{(2E_R m_N)}$. 
It represents the distribution of the weak charge within the nucleus and for small momentum transfer its value is close to unity. 
Having a target nucleus with a high $N$ can increase the cross-section, and if the interaction is coherent, a further enhancement is possible: $\sigma_{CE{\nu}NS} \propto N^2$.

CE$\nu$NS was detected for the first time by the COHERENT collaboration at the Spallation Neutron Source in the U.~S.~A.~\cite{Akimov:2017ade}. 
The difficulty in observing such process was mainly due to the extremely low energy threshold needed and is limited by the detector technology.

This interaction channel is of particular interest to study the neutrino properties since it is equally sensitive to all neutrino flavors. 
Depending on the target nucleus, this process has an interaction cross-section of 3--4 orders of magnitude greater than conventional neutrino detection channels such as inverse beta decay (IBD, see Fig.~\ref{fig:xsec}), and has no energy threshold.
This aspect could have a huge impact in neutrino astronomy where neutrino sources like the sun or supernovae (SN) could be more extensively investigated. This data would be complementary to all other neutrino observatories, which are sensitive to only one neutrino flavor ($\nu_e$/$\overline{\nu}_e$)~\cite{Scholberg:2012id}.

It is clear that a variety of targets can be used for dedicated measurements via CE$\nu$NS, depending on the neutrino energy and on the size of the target nucleus. 
This is particularly true for the cryogenic technique which allows the employment of a wide range of different materials~\cite{Derbin:2014xzr,Casali:2013zzr,Pattavina:2019pxw,Pattavina:2018nhk,Angloher:2016hbv}, unlike semiconductor-based or noble-gaseous based detectors which are limited only to a small number of available materials. 
For this reason, solid state low temperature detectors are ideal devices to be employed for neutrino-nucleus scattering applications.

Thanks to the latest advancements of the low temperature particle detector technology pioneered by CUORE, cryogenic detectors are now suitable devices for the investigation of neutrino properties using CE$\nu$NS interaction. 
A synergistic combination of the unique CUORE cryogenic infrastructure and of the large interaction cross-section of CE$\nu$NS as detection channel can reshape the field of neutrino astronomy. 
A cryogenic detector with the same size as CUORE would be as sensitive as next generation neutrino observatories, such as JUNO~\cite{An:2015jdp}, Hyper-Kamiokande~\cite{Yano:2021usb} or DUNE~\cite{Migenda:2018ljh}, but with a reduced detector volume.
In fact, due to the nature of the interaction and its large cross-section, an overall experimental down-sizing is possible by several orders of magnitude. The potential of such a detector is discussed in details in~\cite{Pattavina:2020cqc}.

The following sections will discuss two specific applications of CE$\nu$NS in a large volume cryogenic detector: the study of solar neutrinos and the investigation of SN events.

\subsubsection{Solar Neutrino Spectroscopy}
Neutrinos are unique messengers for studying the inner core of highly dense astrophysical objects like our Sun or even the Earth.
Thanks to the high interaction cross-section of this detection channel and the low energy threshold achievable with cryogenic detectors, a large array of cryogenic calorimeters can also be operated as a Solar neutrino telescope. CE$\nu$NS, being a flavor blind process, can provide complementary information to all other neutrino observatories, where IBD and other charge-current processes are employed. The information obtained by a flavor-insensitive neutrino detector would enable a precise measurement of the absolute solar neutrino flux.

In the last few decades tremendous effort was spent understanding and measuring neutrino flavor oscillations~\cite{RevModPhys88030502} and developing a robust solar model, the Standard Solar Model (SSM)~\cite{guenther1992standard}. Although the basics of the SSM are established, there are numerous unknowns~\cite{Serenelli:2016dgz} and discrepancies~\cite{PhysRevLett.123.131803}. 
The main challenge in understanding the solar neutrino flux is disentangling neutrino-mixing effects and properties of the neutrino source, which may be greatly addressed by a precise measurement of the absolute neutrino flux.
The relevant open questions are connected to a precise measurement of the $^{8}$B and CNO fluxes, which are directly linked to the solar core temperature and the solar-metallicity problem~\cite{PhysRevC.65.015802}, and to the first measurement of the highest energy component of solar neutrinos, the $\it{hep}$ chain~\cite{Ranucci:2020fcs}. 
%In Fig.~\ref{fig:solar}, we show the expected counting rate for a CUORE-like cryogenic detector  from the higher energy solar neutrino component: $hep$, $^{8}$B and CNO. For this estimation we show three crystal compounds developed for rare events cryogenic investigations: CaWO$_4$~\cite{Abdelhameed:2019hmk}, Ge~\cite{Agnese:2018gze} and TeO$_2$~\cite{Adams:2019jhp}. The plot shows that, if suitable background levels are achieved, the detector will be sensitive to solar neutrino with a relative high rate, depending on the detector energy threshold. 
%As an example, the expected total detector rate for an energy threshold of 10~eV can be as high as 4~eV/tonne/d. \ptsnote{Removed because there was a question about which compound and this line doesn't seem to add much to the information.}

%\begin{figure}[!t]
%\centering
%\includegraphics[width=0.7\textwidth]{images/solar_nu.pdf}
%\caption{CE$\nu$NS energy spectrum from high energy solar neutrinos: $hep$, $^{8}$B and CNO. The colored lines represent the nuclear recoil energy spectra for three different target materials.}
%\label{fig:solar}
%\end{figure}

\subsubsection{Supernova neutrino detection}
A high resolution measurement of the neutrino signal from a future galactic core-collapse SN in a CE$\nu$NS-based cryogenic detector would provide direct empirical evidence of the dynamical and weak-interaction processes that occur in the SN explosion mechanism. Such a measurement will thus test our understanding of stellar core collapse.

The potential of a cryogenic detector for the detection of neutrino from astrophysical sources was  proposed in~\cite{BIASSONI2012151,Pattavina:2020cqc}. In the latter, the potential of the cryogenic technique together with the potential of CE$\nu$NS are reviewed. Precise reconstruction of the temporal evolution of the neutrino emissions can be carried out, revealing whether the core-collapse event lead to the formation of a neutron star or a black hole. Furthermore, the equal sensitivity to all neutrino flavors emitted during the SN event, allows for a highly sensitive detection of the most energetic and most intense components of the neutrino emission: $\nu_{\mu}/\overline\nu_{\mu}$ and $\nu_{\tau}/\overline\nu_{\tau}$. All currently running experiments are mostly sensitive to $\nu_{e}/\overline\nu_{e}$, exploiting IBD and electron scattering as detection channels, thus they have only limited sensitivity to the most relevant SN neutrino emission. 

Following the prescriptions presented in \cite{Pattavina:2020cqc}, we can evaluate the number of events induced in a cryogenic detector by SN neutrinos. A standard core-collapse SN from a progenitor star of $27\ M_\odot$ with equation of state from~\cite{Lattimer:1991nc} will induce a signature in a cryogenic detector made of  1~tonne of TeO$_2$ of tens of events if a detector energy threshold of 100~eV is achieved. This energy threshold can be attained when thermal sensors of the same type employed for direct dark matter searches are implemented (e.g. TESs or MMCs)~\cite{Pirro2017}. %If we now consider the same reference SN event occurring at different distances, we can investigate distance a CUORE-like detector can search for SN events in the neutrino channel. %The probability to detect at least 1 SN neutrino for the reference SN as a function of distance is shown in Fig.~\ref{fig:SN}. A tonne-scale cryogenic detector is able to reach out to the Large Magellanic Cloud for the study of SN event, a similar sensitivity can be only achieved with other technologies (e.g. liquid scintillator) while running detectors with active masses of $\mathcal{O}($>$100~)$~tonne~\cite{Al_Kharusi_2021}.

%\begin{figure}[!t]
%\centering
%\includegraphics[width=0.7\textwidth]{images/SN_nu.pdf}
%\caption{Probability to detect at least 1 SN neutrino event as a function of the SN distance for a reference SN model (see text). A detector energy threshold of 100~eV and a total detector mass of about 1~tonne are assumed}
%\label{fig:SN}
%\end{figure}

%The operation of a large volume cryostat like CUORE can pave the road for the first solid state observatory for extra-galactic neutrinos.

\subsection{Impact on Quantum Technologies}

In the last few decades the development of quantum bits (qubits) through different technologies, such as electron spins, or atomic dipole transitions has seen great progress. The ultimate goal is to develop qubits that remain strongly coupled to each other while remaining highly decoupled from the external world, except during reading, writing, and control operations.

The technique of superconducting circuits has been intensively developed both in the field of particle detectors and of quantum processors. Increases in knowledge of material physics, material interfaces, and microscopic circuits allowed for the development of devices with high fidelity, and fast gate-times that are relatively easy to design and fabricate. The recent demonstration of a system consisting of $\sim$50 qubits~\cite{Martinis2019} also proved the scalability of superconducting circuits, overcoming one of the main limits for most of the competing technologies.
Today, companies and research laboratories are developing transmons, low-impedance flux qubits, but also pioneering ideas such as high impedance flux qubits or hybrid systems. The main challenge in this field is suppressing all the mechanisms leading to the decoherence of single qubit, or to correlated errors in different qubits. 

All the proposed technologies for superconducting qubits would benefit from a cryogenic environment offering high cooling power, low noise, low vibrations, and a shield against electromagnetic interference. 
Recent work proved that also the suppression of radioactivity will play a key role for future technologies. The qubits community has been investigating several sources of decoherence, including dielectric two-level systems~\cite{Burnett2019,Klimov2018}, the absorption of paramagnetic molecules~\cite{deGraaf2018}, Abrikosov vortices trapped in the superconductor~\cite{Wang2014} and quasiparticles~\cite{Catelani2011}. Because of the energy deposition in the superconductor, or on the substrate on which it is deposited, environmental radioactivity can lead to the production of non-equilibrium quasiparticles. 

The study described in Ref.~\cite{Oliver2020} focused on transmons, one of the most popular implementations of superconducting qubits. The authors exposed transmons to a radioactive source, demonstrating that the power released by the source affected its coherence time. This work concluded that environmental radioactivity will be the ultimate limit for transmons, preventing the achievement of coherence times larger than milliseconds. 
Moreover, the work described in Ref.~\cite{Wiler2021} showed that radioactivity induces correlated flips in multiple qubits belonging to the same matrix, undermining the potential of the most popular algorithms for quantum error correction. 
For these reasons, several groups began to develop chips with increased isolation between modules~\cite{gold2021entanglement}, or with ``traps" able to absorb or channel away the energy  produced by radioactive interactions from qubits~\cite{Henriques_2019,martinis2021saving}. 

The complementary study described in Ref.~\cite{Cardani2020} showed that, by operating the superconducting circuit in a cryogenic facility located deep underground, the generation of quasiparticles can be suppressed by more than one order of magnitude, while the dissipation can be decreased by a factor 2-4. 

A cryogenic facility constructed with carefully selected materials and heavily shielded from environmental radioactivity could further extend this result, offering an attractive setting for the operation of next generation quantum computers.

\section{Conclusions}\label{sec:conclusion}
Large scale cryogenic facilities are increasingly becoming important in diverse fields of science from dark matter detection to operation of next generation quantum processors. Although the exact specifications vary based on the field of study, there are several shared characteristics. The requirements for large-scale cryogenic facilities typically include high cooling power, low noise, and low radioactive backgrounds. The CUORE experiment has built and has been operating a cryogenic facility that satisfies the above requirements to perform a high sensitivity search for \zeronu.

% In this article, a detailed description of the design of the CUORE cryogenic infrastructure motivated by the physics requirements of the experiment was given. The commissioning and performance of the CUORE cryostat along with the difficulties faced and their solutions were discussed. Potential future improvements for the CUORE cryostat design and operation were summarized. 

The unprecedented success in the realization and long-term operation of the CUORE cryostat has long-lasting implications for the experiments in particle and nuclear physics and dark matter detection as well as in quantum computing. 
Upcoming large scale cryogenic facilities would immensely benefit from the lessons learned by the CUORE collaboration from the design phase all the way to the commissioning and operation of the CUORE cryogenic facility.
The next generation cryogenic experiments like CUPID can profit from the low backgrounds and good energy resolution demonstrated by the CUORE experiment and further improve sensitivity to \zeronu~by implementing dual readout for background reduction and reducing the resolution to a greater extent based on the experiences from CUORE.
Large cryogenic facillities, such as what the CUORE cryostat provides, offer a promising way to detect astrophysical neutrinos. These can act as unique probes to study unexplored astrophysics such as the core of the sun and the interaction processes involved in supernovae.
The ability to achieve low backgrounds using a wide choice of materials with a capability to perform event-by-event background rejection also makes large-scale cryogenic infrastructures a favorable choice for future DM experiments.
The features such as high cooling power, low thermal and vibrational noise, and reduced radioactive and electromagnetic backgrounds provided by a cryogenic facility will also be quite valuable to the emerging, and highly competitive, field of quantum information sciences.
%Some near-term use cases for tonne-scale cryogenic facilities like CUORE were outlined. 

\section*{Acknowledgements}
The CUORE Collaboration thanks the directors and staff of the Laboratori Nazionali del Gran Sasso and the technical staff of our laboratories. 
This work was supported by the Istituto Nazionale di Fisica Nucleare (INFN); the National Science Foundation under Grant Nos. NSF-PHY-0605119, NSF-PHY-0500337, NSF-PHY-0855314, NSF-PHY-0902171, NSF-PHY-0969852, NSF-PHY-1307204, NSF-PHY-1314881, NSF-PHY-1401832, and NSF-PHY-1913374; and Yale University. 
This material is also based upon work supported by the US Department of Energy (DOE) Office of Science under Contract Nos. DE-AC02-05CH11231 and DE-AC52-07NA27344; by the DOE Office of Science, Office of Nuclear Physics under Contract Nos. DE-FG02-08ER41551, DE-FG03-00ER41138, DE-SC0012654, DE-SC0020423, DE-SC0019316; and by the EU Horizon2020 research and innovation program under the Marie Sklodowska-Curie Grant Agreement No. 754496.  
This research used resources of the National Energy Research Scientific Computing Center (NERSC).
This work makes use of both the DIANA data analysis and APOLLO data acquisition software packages, which were developed by the CUORICINO, CUORE, LUCIFER and CUPID-0 Collaborations. 

\bibliography{cuore-ppnp}

\end{document}

%% file: cuore-authors.tex
% Updated on Friday, 16 Aug 2021 + PRL 2018 + Tech 
\author[USC]{D.~Q.~Adams}
\author[USC]{C.~Alduino}
\author[INFNMilano]{F.~Alessandria\fnref{fn2}}
\author[UCLA]{K.~Alfonso}
\author[Como,INFNMiB]{E.~Andreotti}
\author[USC]{F.~T.~Avignone~III}
\author[INFNLegnaro]{O.~Azzolini}
\author[LNGS]{M.~Balata}
\author[USC]{I.~Bandac}
\author[BerkeleyPhys,LBNLNucSci]{T.~I.~Banks}
% ...

\address[USC]{Department of Physics and Astronomy, University of South Carolina, Columbia, SC 29208, USA}
\address[INFNMilano]{INFN -- Sezione di Milano, Milano I-20133, Italy}
\address[UCLA]{Department of Physics and Astronomy, University of California, Los Angeles, CA 90095, USA}
\address[Como]{Dipartimento di Fisica e Matematica, Universit\`{a} dell’Insubria, Como I-22100, Italy}
\address[INFNMiB]{INFN -- Sezione di Milano Bicocca, Milano I-20126, Italy}
% ...
\address[INFNLegnaro]{INFN -- Laboratori Nazionali di Legnaro, Legnaro (Padova) I-35020, Italy}
% ...
\address[LNGS]{INFN -- Laboratori Nazionali del Gran Sasso, Assergi (L'Aquila) I-67100, Italy}
\address[BerkeleyPhys]{Department of Physics, University of California, Berkeley, CA 94720, USA}
% ...
\address[LBNLNucSci]{Nuclear Science Division, Lawrence Berkeley National Laboratory, Berkeley, CA 94720, USA}
% ...

\fntext[fn1]{Deceased}
\fntext[fn2]{Retired}